\documentclass[
    aps,pre,twocolumn,
    longbibliography
    reprint,
    superscriptaddress,
    floatfix
]{revtex4-1}

\usepackage[utf8]{inputenc}

\usepackage{amsmath,amssymb}
\usepackage{graphicx}

\usepackage{siunitx}

\usepackage{xcolor}
\definecolor{darkred}{RGB}{200,0,0}
\definecolor{myTeal}{RGB}{0,100,150}

\definecolor{linkColor}{RGB}{0,70,120}

\usepackage[colorlinks=true, allcolors=linkColor,pdfborder={0 0 0},pdfencoding = auto]{hyperref}

\usepackage[normalem]{ulem}

\DeclareMathOperator{\diag}{diag}

\usepackage{bm}
\def\fvec{\bm{f}}
\def\uvec{\mathbf{u}}
\def\cvec{\mathbf{c}}
\def\mvec{\mathbf{m}}
\def\nvec{\mathbf{n}}

\newcommand\kD{k_\mathrm{D}} 
\newcommand\kdD{k_\mathrm{dD}}
\newcommand\kde{k_\mathrm{de}}
\newcommand\kdE{k_\mathrm{dE}}

\newcommand\cD{c_\mathrm{D}}
\newcommand\cDD{c_\mathrm{DD}}
\newcommand\cDT{c_\mathrm{DT}}
\newcommand\cE{c_\mathrm{E}}
\newcommand\md{m_\mathrm{d}}
\newcommand\mde{m_\mathrm{de}}

\newcommand\nD{n_{\rule{0pt}{1.3ex} \mathrm{D}}}
\newcommand\nE{n_{\rule{0pt}{1.3ex} \mathrm{E}}}

\newcommand\DD{D_\mathrm{D}}
\newcommand\Dd{D_\mathrm{d}}
\newcommand\Dde{D_\mathrm{de}}
\newcommand\DE{D_\mathrm{E}}

\def\dnD{\Delta n_{\rule{0pt}{1.3ex} \mathrm{D}}}
\def\dnE{\Delta n_{\rule{0pt}{1.3ex} \mathrm{E}}}
\def\dnDE{\Delta n_{\rule{0pt}{1.3ex} \mathrm{D,E}}}

\def\dcD{\Delta c_\mathrm{D}^{*}}
\def\dcE{\Delta c_\mathrm{E}^{*}}

\def\dmd{\Delta m_\mathrm{d}^{*}}
\def\dmde{\Delta m_\mathrm{de}^{*}}

\def\DhD{\mathcal{D}_\mathrm{D}^{}}
\def\DhE{\mathcal{D}_\mathrm{E}^{}}
\def\Dhd{\mathcal{D}_\mathrm{d}^{}}
\def\Dhde{\mathcal{D}_\mathrm{de}^{}}

\begin{document}

\title{Diffusive coupling of two well-mixed compartments elucidates elementary principles of protein-based pattern formation}

\author{Fridtjof Brauns} 
\affiliation{Arnold Sommerfeld Center for Theoretical Physics and Center for NanoScience, Department of Physics, Ludwig-Maximilians-Universit\"at M\"unchen, Theresienstra\ss e 37, D-80333 M\"unchen, Germany}
\author{Jacob Halatek}
\affiliation{Biological Computation Group, Microsoft Research, Cambridge CB1 2FB, UK}
\author{Erwin Frey}
\email{frey@lmu.de}
\affiliation{Arnold Sommerfeld Center for Theoretical Physics and Center for NanoScience, Department of Physics, Ludwig-Maximilians-Universit\"at M\"unchen, Theresienstra\ss e 37, D-80333 M\"unchen, Germany}

\begin{abstract}
    Spatial organization of proteins in cells is important for many biological functions. In general, the nonlinear, spatially coupled models for protein-pattern formation are only accessible to numerical simulations, which has limited insight into the general underlying principles. To overcome this limitation, we adopt the setting of two diffusively coupled, well-mixed compartments that  represents the elementary feature of any pattern---an interface. For intracellular systems, the total numbers of proteins are conserved on the relevant timescale of pattern formation. Thus, the essential dynamics is the redistribution of the globally conserved mass densities between the two compartments. 
    We present a phase-portrait analysis in the phase-space of the redistributed masses that provides insights on the physical mechanisms underlying pattern formation.
    We demonstrate this approach for several paradigmatic model systems. In particular, we show that the pole-to-pole Min oscillations in \textit{Escherichia coli} are relaxation oscillations of the MinD polarity orientation. This reveals a close relation between cell polarity oscillatory patterns in cells. Critically, our findings suggest that the design principles of intracellular pattern formation are found in characteristic features in these phase portraits (nullclines and fixed points). These features are not uniquely determined by the topology of the protein-interaction network but depend on parameters (kinetic rates, diffusion constants) and distinct networks can give rise to equivalent phase portrait features. 
\end{abstract}

\maketitle

\section{Introduction}

The spatial intracellular organization of proteins by reactions (protein-protein interactions) and diffusion has received growing attention in recent years; for recent reviews see Refs.~\cite{Goryachev.Leda2017, Rappel.Edelstein-Keshet2017, Chiou.etal2017, Lang.Munro2017, Frey.etal2018, Halatek.etal2018, Murray.Howard2019, Cheng.etal2020}.
Gaining intuition and theoretical insight into the spatiotemporal protein dynamics remains challenging owing to the complexity arising from the spatial coupling and nonlinear reaction terms.
Therefore, insights often remain restricted to specific mathematical models.
A systematic understanding is hard to achieve, in particular if there are multiple protein species with several conformational states involved (complex interaction network).
Thus, finding the elementary principles underpinning protein-based pattern formation still remains a largely open question.

\begin{figure}
    \centering
    \includegraphics{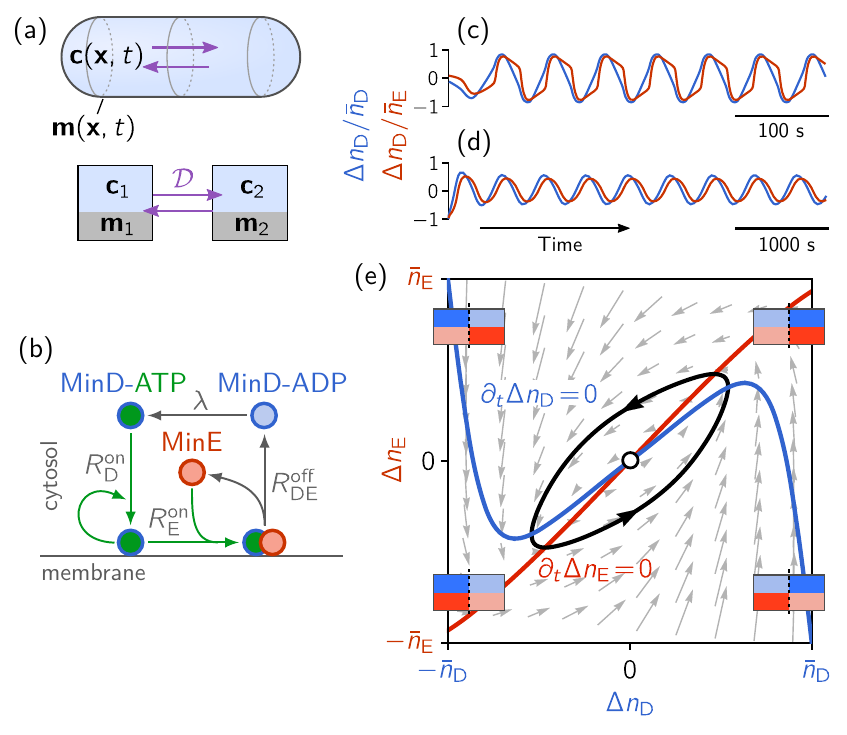}
    \caption{%
      Reduction from the full dynamics in cell-geometry to the phase portrait of  mass-redistribution dynamics.
      (a) Spherocylinder geometry of a rod-shaped \textit{E.\ coli} cell (top) and reduced two-compartment setting representing the two cell halves (bottom). Purple arrows illustrate diffusive mass transport.
      (b) Protein-interaction network of the Min system of \textit{E.\ coli} (see main text Sec.~\ref{sec:Min} for details.
      (c) Time traces of the protein mass in the compartments relative to the mean, $\Delta n_\mathrm{D,E}$ defined via $n_\mathrm{D,E}^{(1,2)} = \bar{n}_\mathrm{D,E} \pm \Delta n_\mathrm{D,E}$, showing the pole-to-pole oscillations in three-dimensional cell geometry.
      (d) Oscillations persist in the two-compartment setting, with diffusive exchange rates set to a slow time scale.
      (e) Phase portrait of the mass-redistribution dynamics showing the flow field (gray arrows) and the nullclines of MinD and MinE redistribution (blue and red lines). The origin $(0,0)$ corresponds to the homogeneous state which is unstable against perturbations redistributing mass. The limit cycle trajectory (black) corresponds to pole-to-pole oscillations. The cartoons in the four corners illustrate the two-compartments (separated by a vertical dashed line) where the color intensity indicates the mass distribution of MinD (blue) and MinE (red) in the respective quadrant of the phase portrait.
    }
    \label{fig:motivation}
\end{figure}

To simplify the analysis on a technical level, systems of two diffusively coupled, well-mixed compartments (also called `boxes', `reactors', `cells', or `patches') have been widely used in earlier literature.
In fact Turing himself used the setting of diffusively coupled compartments (called ``cells'')  in his pioneering work to show that diffusion can destabilize otherwise stable reactions, thus leading to spatial pattern formation \cite{Turing1952}.
Physically, the two-compartment setting represents the elementary feature of any pattern---an interface connecting a low density region to a high density region.
In the context of intracellular pattern formation, the two compartments typically represent the polar zones of rod-shaped cells, such as \textit{E.~coli} bacteria (see Fig.~\ref{fig:motivation}a), \textit{M.~xanthus} bacteria \cite{Guzzo.etal2018,Tostevin.etal2021}, and fission yeast (\textit{S.~pombe}) \cite{Das.etal2012, Xu.Jilkine2018}.

In a broader context, two-compartment systems also have been realized in experiments, using diffusively coupled CSTRs (continuously stirred tank reactors) \cite{Bar-Eli.Geiseler1981} and recently using nanometer scale microfluidic devices \cite{Litschel.etal2018a, Norton.etal2019}.
Furthermore, in population dynamics, they are known as ``two-patch systems'' and have been used to study toe role of spatial coupling and patterning in ecology, see e.g.\ \cite{Holt1985,Blasius.etal1999,Salomon.etal2010}.

In this manuscript, we focus on protein-based pattern formation in cells.
A key property of such intracellular pattern formation is that the total number of proteins is conserved on the relevant time scale of pattern formation \cite{Goryachev.Pokhilko2008, Goehring.etal2011, Halatek.Frey2012, Trong.etal2014, Halatek.etal2018}.
Recent works \cite{Halatek.Frey2018, Brauns.etal2020b} suggest that (diffusive) mass redistribution is the key physical process driving pattern formation in mass-conserving reaction--diffusion systems.
Based on this insight, a framework termed \emph{local equilibria theory} has been developed \cite{Brauns.etal2020b}.
The basic idea of this framework is to consider the system as decomposed into (notional) compartments, small enough to be effectively well-mixed. Within each compartment, the reactive dynamics conserves the mass(es). The reactive equilibria (steady states) of the reactions within an isolated compartment, controlled by these local masses, serve as proxies for the local dynamics. Diffusive coupling of the compartments redistributes masses between them. In turn, the changing local masses shift the local reactive equilibria and potentially change their stability.
Thinking about reaction--diffusion systems in terms of this interplay between mass-redistribution and shifting local equilibria has proven a powerful approach to study their complex nonlinear dynamics \cite{Halatek.Frey2018,Brauns.etal2020b,Brauns.etal2020a,Brauns.etal2020,Wigbers.etal2020}. 

Here we adopt the two-compartment setting and show how this way of thinking can be made explicit in the form of simple graphical constructions and a phase portrait analysis in the phase space of the redistributed masses.
This will enable us to gain insights on the physical mechanisms underlying pattern formation that would otherwise remain hidden.
Importantly, and in contrast to previous works \cite{Mori.etal2008, Holmes.etal2015, Diegmiller.etal2018, Guzzo.etal2018, Tostevin.etal2021}, we do \emph{not} assume the fast diffusing (cytosolic) components to be well mixed.
In other words, we explicitly allow for cytosolic gradients between the two compartments. As we will see later, this is important understand the physical mechanisms underlying pattern formation. In particular, it is key to explain the pole-to-pole oscillations of Min proteins in \textit{E.\ coli}.

\paragraph{Motivation.}

Let us present the main motivation for this work using the pole-to-pole oscillations of Min proteins in \textit{E.\ coli} as a concrete example without going into technical details (which will be presented below). 
Put briefly, the pole-to-pole oscillations are driven by two types of proteins, MinD and MinE, which cycle between membrane-bound and cytosolic states and interact with each other on the membrane (Fig.~\ref{fig:motivation}b), while the total masses of MinD and MinE ($
\nD$ and $\nE$) remain conserved.
A key insight from previous works is that spatial redistribution of such globally conserved masses constitutes the essential degrees of freedom of mass-conserving reaction diffusion systems \cite{Halatek.Frey2018}. Indeed, mapping the Min system to the two-compartment setting and tuning the diffusive exchange rates to a slow time scale retains the qualitative features of the pole-to-pole oscillations (Fig.~\ref{fig:motivation}c,d). On the slow time scale, only the 
masses in the two compartments $n_\mathrm{D,E}^{(1,2)}$ remain as dynamic variables.
Because of mass conservation, the average masses $\bar{n}_\mathrm{D,E}$ remain constant. Defining the redistributed masses, $\dnD$ and $\dnE$, via $n_\mathrm{D,E}^{(1,2)} = \bar{n}^{}_\mathrm{D,E} \pm \dnDE$, 
we can visualize the dynamics in the two-dimensional $(\dnD,\dnE)$-phase space, Fig.~\ref{fig:motivation}e, where we plot the flow field and its nullclines. 
Along the nullclines the rate of mass-exchange between the compartments vanishes. We hence refer to them as mass-redistribution nullclines. 
The phase portrait shows the characteristics of relaxation oscillations.
In this paper, we show that the Min pole-to-pole oscillations are indeed \emph{spatial} relaxation oscillations of the MinD polarity orientation.

This example shows how important qualitative features of mass-conserving reaction--diffusion (MCRD) systems can be obtained from a phase portrait analysis in the phase space of the redistributed masses.
In the following, we show how this phase portrait can be constructed systematically, starting from the reaction--diffusion equations.
We show what determines the structure of the phase space flow and derive a simple geometric relation between the \emph{mass-redistribution} nullclines and the \emph{reactive} nullclines of the local reaction kinetics.

\paragraph{Structure of the paper.}
To introduce the basic elements of our analysis, we first study MCRD systems with two components, e.g.\ the membrane-bound and cytosolic state of a single protein species (see Sec.~\ref{sec:two-components}).
We then generalize the nullcline-based approach to systematically derive the phase portrait of the Min system of \textit{E.\ coli} shown in Fig.~\ref{fig:motivation}e. This construction then allows us to study the role of diffusive mass redistribution of MinD and MinE for the formation of Min-protein patterns.
Finally, we apply the same approach to two other paradigmatic model systems: PAR polarity of \textit{C.\ elegans} and Cdc42 polarity of budding yeast.
Comparing the different nullcline geometries of these systems allows one to classify their pattern-forming mechanisms (see Sec.~\ref{sec:PAR-and-Cdc42}).
Such a nullcline-based classification provides intuition for the role of various elements in the biochemical network. Moreover, it might guide model building and serve as a first step of analysis for systems that are biochemically not as well characterized as the aforementioned examples. 
In the Conclusions, Sec.~\ref{sec:conclusions}, we discuss important implications of our work, both specific to the Min system and in a broader context, and give an outlook to promising future research directions.

\section{Two-component MCRD systems}
\label{sec:two-components}

Two-component MCRD systems have been previously used as conceptual models for cell polarity~\cite{Otsuji.etal2007,Goryachev.Pokhilko2008,Mori.etal2008,Chiou.etal2018}.
In this section, we apply local equilibria theory \cite{Brauns.etal2020b,Halatek.Frey2018} to these systems in the two-compartment setting. 
In this simplified setting, the formulation of local equilibria theory is technically simpler than in spatially continuous systems
\footnote{A detailed analysis of two-component MCRD systems in a spatially continuous setting can be found in Ref.~\cite{Brauns.etal2020b}.}.
Importantly, the approach developed below for two-component MCRD systems can be generalized to systems with more components and more conserved masses such as those studied in Sections~\ref{sec:Min} and~\ref{sec:PAR-and-Cdc42}, where the new approach yields novel insights.

Let us denote the concentrations of the two components in compartment $i \in \{1,2\}$ by $\uvec_i = (m_i,c_i)$, where $m_i$ and $c_i$ are the concentration of membrane-bound and cytosolic proteins, respectively.
The reaction kinetics $\fvec = (f,-f)$ within each compartment account for the attachment and detachment to and from the membrane. Importantly, they conserve the local total density (mass) $n_i = m_i + c_i$ in each of the two compartments individually. 
Mass is transferred between the compartments by a diffusive exchange process that acts to even out concentration differences. 
Denoting the diffusive exchange rates in the matrix $\mathfrak{D} = \diag (\mathcal{D}_m, \mathcal{D}_c)$, we have the coupled compartment dynamics in vector notation
\begin{equation} 
\label{eq:two-compartment-dyn}
\begin{split}
	\partial_t \uvec_1 
	&= \mathfrak{D} \big(\uvec_2- \uvec_1\big) + \fvec \big(\uvec_1\big), \\
	\partial_t \uvec_2 
	&= \mathfrak{D} \big(\uvec_1 - \uvec_2\big) + \fvec \big(\uvec_2\big).
\end{split}
\end{equation}
Since both the local reactions and the diffusive exchange are mass conserving, the average total density $\bar{n} = (n_1 + n_2)/2$ is a constant of motion.
In Appendix~\ref{app:LSA-continuous}, it is shown how the (diffusive) exchange rates $\mathcal{D}_{m,c}$ can be related to the diffusion constants $D_{m,c}$ in a spatially continuous system, in such a way that the linearized dynamics of Eq.~\eqref{eq:two-compartment-dyn} near a homogeneous steady state is identical to the linearized dynamics of a single Fourier mode ${\sim} \cos(\pi x/L)$ in the spatially continuous system on the interval $[0,L]$ with no-flux boundary conditions. 
For patterns with large amplitudes, nonlinearities lead to mode coupling in a spatially continuous system. 
This is not captured by the two-component system which only describes the dynamics at a single length scale. Nonetheless, one can gain a good qualitative understanding of the full nonlinear pattern formation process, including the termination of the pattern-forming instability in a stationary pattern.

\subsection{Setting the stage: phase-space geometry of two-component MCRD systems}

In the following, we present the key concepts of local equilibria theory in the two-compartment setting.
Because of mass conservation, only the mass density difference with respect to the mean $\Delta n := (n_1 - n_2)/2$ is a dynamic variable, while $\bar{n}$ is a \emph{control parameter}. Thus, we can rewrite the local masses as $n^{}_{1,2}(t) = \bar{n} \pm \Delta n(t)$.
Adding the equations for $\partial_t m_1$ and $\partial_t c_1$ (Eq.~\eqref{eq:two-compartment-dyn} yields $\partial_t n_1$, and thus
\begin{equation} \label{eq:mass-dyn-2c}
	\partial_t \Delta n(t) = -\mathcal{D}_m \Delta m - \mathcal{D}_c \Delta c,
\end{equation}
where $\Delta \uvec = (\Delta m, \Delta c):= \uvec_1 - \uvec_2$ and we used that $\partial_t \bar{n} = 0$.
Observe that the reaction terms cancel because they conserve the mass in each compartment individually. 
Thus, the dynamics of the total density is solely determined by concentration differences in $m$ and $c$ between the two compartments. These concentration differences approximate the gradients in the spatially continuous system.

\begin{figure*}
	\centering
	\includegraphics{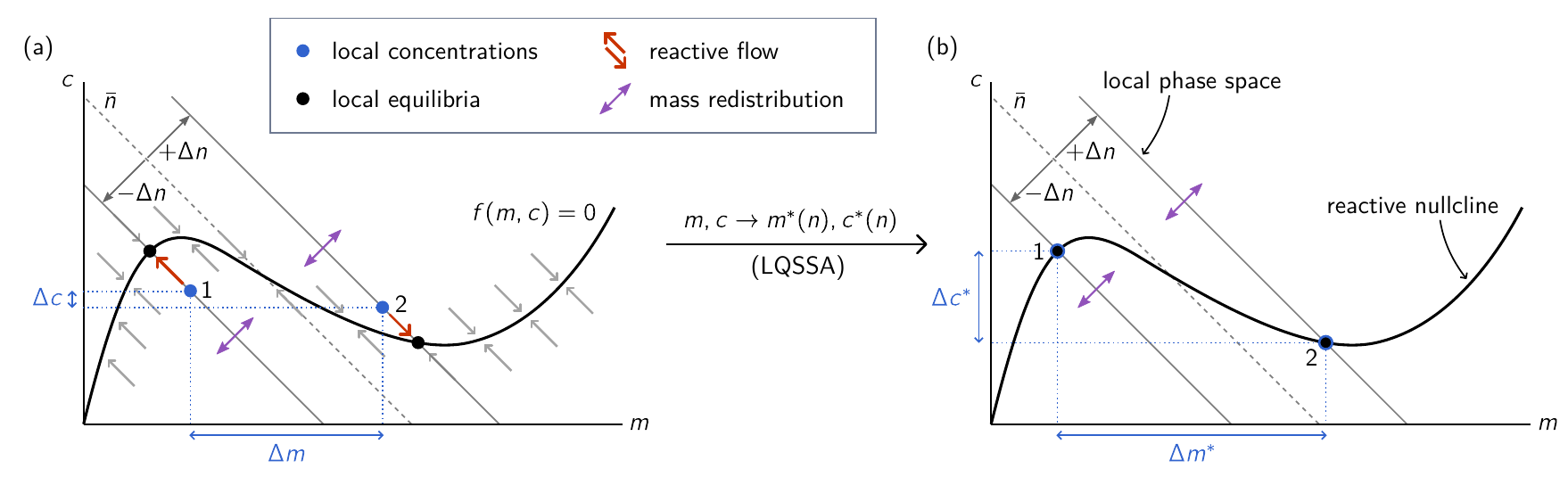}
	\caption{%
	Phase space structure of a two-compartment two-component MCRD system, with reaction and diffusion on the same time scale (a) and with diffusion set to a slower time scale (b).
	The concentrations $(m_i,c_i)$ in the two compartments are marked by blue dots, labelled 1 and 2, respectively. 
	The local phase spaces corresponding to the masses in the two compartments $n_{1,2} = \bar{n} \pm \Delta n$ are shown as gray lines.
	Gray arrows indicate the reactive flow towards the reactive nullcline $f = 0$ (solid black line).
	Black dots mark the local equilibria (intersection points between reactive nullcline and local phase spaces) and red arrows indicate the reactive flow towards these local equilibria.
	(b) When diffusion is set to a slower time scale, the local concentrations adiabatically follow the reactive nullcline. Thus, the only remaining degree of freedom is the mass difference $\Delta n$, whose dynamics is governed by the concentration differences $\Delta m^*(\Delta n)$ and $\Delta c^*(\Delta n)$ (see Eq.~\eqref{eq:mass-dyn-2c}).
	}
	\label{fig:2cMcRD_phase-space}
\end{figure*}

To understand how these concentration differences are governed by the reaction kinetics, consider the $(m,c)$-phase plane of the reaction kinetics  (see Fig.~\ref{fig:2cMcRD_phase-space}a). 
While this phase plane is two-dimensional, mass conservation also implies that reactive flow $(f,-f)$ in each compartment $i$ is constrained to a respective linear subspace $m_i + c_i = n_i$.
We term these subspaces the \emph{local phase spaces} of each compartment \cite{Halatek.Frey2012,Brauns.etal2020b}.
Here, and in the following, the term \emph{local} always refers to the properties of a single (notionally isolated) compartment.
Correspondingly, we define as \emph{local reactive equilibrium} the point within the local phase space where the reaction kinetics are balanced, i.e.\ where the reactive flow vanishes ($f=0$):
\begin{equation} \label{eq:loc-eq}
	\uvec^*(n_i) \; : \;
	\left\{ \begin{array}{rl}
	    \fvec(\uvec^*) =& \!\! 0, \\
	    m^* + c^* =& \!\! n_i.
	\end{array} \right.
\end{equation}
Geometrically, the local equilibria are the intersection points between the local phase spaces and the reactive nullcline (see Fig.~\ref{fig:2cMcRD_phase-space}) 
\footnote{Depending on the nullcline shape, there can be several reactive equilibria for a given total density~\cite{Brauns.etal2020b}. We focus here on the case where only one, stable equilibrium exists.}.
These local equilibria determine the steady state (reactive equilibrium) in each compartment that would be reached if, given the local masses $n_1$ and $n_2$, the two compartments were isolated, i.e.\ if the diffusive exchange between the compartments was shut off.
Thus, the local equilibria serve as proxies for the local reactive flow within each of the compartments (red arrows in Fig.~\ref{fig:2cMcRD_phase-space}a).

Diffusive coupling between the compartments redistributes mass between the compartments. This is reflected in the shifting of the local phase spaces \emph{in the $(m,c)$-phase plane}, as indicated by the purple arrows in Fig.~\ref{fig:2cMcRD_phase-space}a,b).
As a result, the local reaction kinetics change since the local equilibria move in the $(m,c)$-phase plane.
In the following we will elucidate this interplay between diffusive mass-redistribution and shifting local equilibria in the most elementary form.

\subsection{Limit of slow mass exchange}

To separate the roles of local reactions and diffusive mass redistribution, we consider a situation where the latter occurs on a much slower time scale than the former
\footnote{
More specifically, this means $\mathcal{D}_{m,c} \ll \sigma_\mathrm{loc}$, where $\sigma_\mathrm{loc}$ is the eigenvalue of the linearized local reaction kinetics with the smallest absolute real part. In the case of a two-component system, $\sigma_\mathrm{loc} = f_c - f_m$, with $f_\alpha = \partial_\alpha f|_{\uvec^*}$.}.
In this limit, the cytosolic and membrane concentrations in each compartment adiabatically follow the local equilibria that depend on the local masses $n_i$, as encoded by the shape of the reactive nullcline in the $(m,c)$-phase plane (see Fig.~\ref{fig:2cMcRD_phase-space}b).
We can, therefore, approximate the densities by their respective equilibrium values
\begin{equation} \label{eq:LQSSA}
	\uvec_i(t) \approx \uvec^*\big(n_i(t)\big).
\end{equation}
We term this the \emph{local quasi-steady state approximation} (LQSSA).
The dynamics of the mass difference $\Delta n$ is then governed by a closed equation
\begin{equation}
	\partial_t \Delta n(t) \approx -\mathcal{D}_m \Delta m^*(\Delta n) - \mathcal{D}_c \Delta c^*(\Delta n),
\end{equation}
with the shorthand notation for the concentration differences between the two compartments:
\begin{equation} \label{eq:scaffolded-gradients}
	\Delta \uvec^*(\Delta n) := \uvec^*(\bar{n} + \Delta n) - \uvec^*(\bar{n} - \Delta n).
\end{equation}
In this approximation, the roles of local reactive dynamics and diffusive mass exchange are clearly separated.
The concentrations only change if the local phase spaces shift due to mass redistribution.
In turn, the mass fluxes from one compartment to the other are determined by the concentration gradients $\Delta \uvec^*(\Delta n)$, weighted by the respective exchange rates $\mathcal{D}_{m,c}$.
This nonlinear feedback between shifting equilibria and mass redistribution is the basic mechanism underlying pattern formation in mass-conserving reaction diffusion systems \cite{Halatek.Frey2018, Brauns.etal2020b}.
Importantly, the role of the reaction kinetics is fully encoded in the shape of the reactive nullcline, i.e.\ the functional dependence of the reactive equilibrium concentrations $\uvec^*(n)$ on the total density $n$.

The local masses $n_i$ within each compartment play the role of \emph{control variables} \cite{Halatek.Frey2018} that determine the position of the local phase spaces (and thus the position of the reactive equilibria) within the $(m,c)$-phase plane.
At the same time the local masses are also dynamic variables that change by means of diffusive mass redistribution between the compartments.
Accordingly, we refer to the phase space of the redistributed masses as \emph{control space}.
In the two-component MCRD system, the only control variable is the mass difference $\Delta n$, such that the control space is one-dimensional.

Typically, diffusion on the membrane is orders of magnitude slower than in the cytosol, $D_m \ll D_c$ such that its contribution to mass redistribution can be neglected; see e.g.\ Refs.~\cite{Meacci.etal2006,Bendezu.etal2015}.  
Hence, to simplify the following analysis, we neglect the slow membrane diffusion (i.e.\ we set $D_m = 0$), such that
\begin{equation} \label{eq:mass-dyn_cyt-only}
\begin{aligned}
	\partial_t \Delta n(t) &= -\mathcal{D}_c \Delta c^*(\Delta n) \\
	&= -\mathcal{D}_c \big[ c^*(\bar{n} + \Delta n) - c^*(\bar{n} - \Delta n) \big].
\end{aligned}
\end{equation}
Generalization to account for the effect of membrane diffusion is straightforward by changing variables from $c$ to the `mass-redistribution potential' $\eta := c + (D_m/D_c) m$ \cite{Bracha.etal2018}.

\begin{figure}[tbp]
	\centering
	\includegraphics{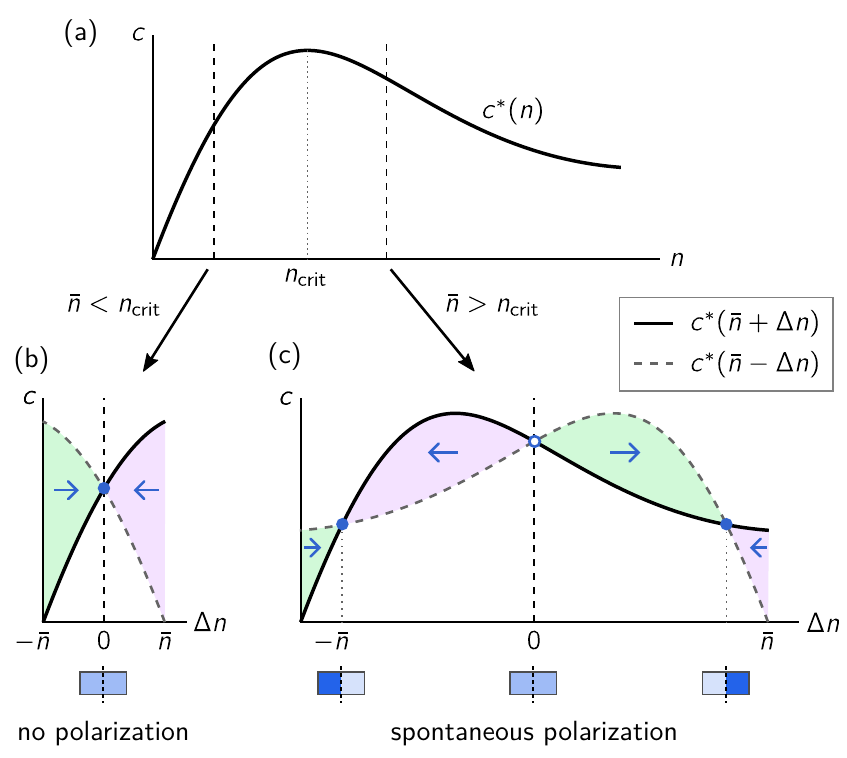}
	\caption{%
	Graphical construction of the control-space dynamics for two-component MCRD systems.
	(a) Reactive nullcline (line of reactive equilibria) $c^*(n)$.
	(b) For $\bar{n} < n_\mathrm{crit}$, the lines $c^*(\bar{n} + \Delta n)$ (black solid line) and $c^*(\bar{n} - \Delta n)$ (gray dashed line) only intersect once at $\Delta n = 0$, corresponding to the homogeneous steady state. The flow direction in control space (indicated by blue arrows) is determined by the sign of the difference between $c^*(\bar{n} + \Delta n)$ and $c^*(\bar{n} + \Delta n)$, as indicated by the green and purple shading; cf.\ Eq.~\eqref{eq:mass-dyn_cyt-only}.
	(c) For $\bar{n} > n_\mathrm{crit}$, there are two additional intersection points between $c^*(\bar{n} + \Delta n)$ and $c^*(\bar{n} - \Delta n)$, corresponding polarized steady states. The flow in control space is directed away from the homogeneous state $\Delta n = 0$, which is therefore unstable, and drives the system towards one of the stable polarized states.
	}
	\label{fig:2cMcRD-NC-construction}
\end{figure}

Equation~\eqref{eq:mass-dyn_cyt-only}, has a simple geometric interpretation as shown in Fig.~\ref{fig:2cMcRD-NC-construction}b,c. The term in the brackets in Eq.~\eqref{eq:mass-dyn_cyt-only} expresses the difference between the nullcline (solid, black line) and its mirror image (dashed gray line)  reflected at the point $\bar{n}$.
Depending on the nullcline slope at $\bar{n}$, the resulting dynamics $\partial_t \Delta n$, indicated by the blue arrows, is qualitatively different.
For a positive slope, $\partial_n c^*(n)|_{\bar{n}} > 0$, following a small perturbation from the ``homogeneous'' state $\Delta n = 0$ the system returns to the $\Delta n = 0$; see Fig.~\ref{fig:2cMcRD-NC-construction}b.
In contrast, for a negative slope, $\partial_n c^*(n)|_{\bar{n}} < 0$, the homogeneous state is unstable; see Fig.~\ref{fig:2cMcRD-NC-construction}c.
This criterion for a lateral instability (instability against spatially inhomogeneous perturbations) was previously derived in Ref.~\cite{Brauns.etal2020b} for spatially continuous systems.
The physical mechanism for this \emph{mass-redistribution instability} is that the reactive equilibrium shifts to lower concentration of the fast diffusing (cytosolic) component, $c^*(n)$, when the total density $n$ is increased, and vice versa. 
Hence, a small perturbation $\delta n$ results in a gradient $\Delta c$ that transports mass from the compartment with lower mass to the compartment with higher mass. 
This amplification mechanism drives the instability.

The growth of the mass difference $\Delta n$ will stop once the cytosolic gradient $\Delta c^*(\Delta n)$ vanishes, i.e.\ when the cytosolic concentration is the same in both compartments,
$c^*(\bar{n} + \Delta n) = c^*(\bar{n} - \Delta n)$.
Thus, stationary states can be determined graphically as the intersection points of the nullcline $c^*(n)$ with its own mirror image, mirrored at $\bar{n}$, as illustrated in Fig.~\ref{fig:2cMcRD-NC-construction}c.
The intersection point at $\Delta n = 0$ always exists by construction, and corresponds to the homogeneous steady state.
The two intersection points at $\Delta n \neq 0$ represent polarized steady states.

In summary, we have shown how one can graphically construct the mass-redistribution dynamics of two-compartment systems with one conserved mass simply based on the reactive nullcline $\mathbf{u}^*(n)$.
In the next section, we will generalize this construction to systems with two conserved masses.

\section{Two-conserved masses: the example of Min-protein oscillations}
\label{sec:Min}

The Min-protein system is a paradigmatic model system for intracellular pattern formation. It was discovered in \textit{E.\ coli}, where the pole-to-pole oscillations of the proteins MinD and MinE allow the cell to position its division machinery at midcell \cite{Lutkenhaus2007,Raskin.deBoer1999}.
This spatial oscillation, i.e.\ the alternating accumulation of the proteins at the two cell poles is driven by cycling of MinD and MinE between cytosolic and membrane bound states, fuelled by ATP (details described below).
Subsequent to its reconstitution \textit{in vitro} \cite{Loose.etal2008}, the Min system has been studied in great detail, both experimentally \cite{Loose.etal2008,Ivanov.Mizuuchi2010,Zieske.Schwille2013,Zieske.Schwille2014,Vecchiarelli.etal2014,Zieske.etal2016,Caspi.Dekker2016,Denk.etal2018,Kohyama.etal2019,Godino.etal2019,Glock.etal2019a,Glock.etal2019,Brauns.etal2020a} and theoretically \cite{Howard.etal2001,Kruse2002,Huang.etal2003,Halatek.Frey2012,Wu.etal2016,Halatek.Frey2018,Denk.etal2018,Glock.etal2019a,Brauns.etal2020a}. 
This research has revealed a bewildering zoo of patterns, including traveling waves, standing waves, spatiotemporal chaos, and defect mediated turbulence, observed in different experimental setups (including microfluidic devices \cite{Vecchiarelli.etal2014,Brauns.etal2020a} and vesicles \cite{Kohyama.etal2019,Godino.etal2019}). 
Recent works employing local-equilibria theory to interpret data from numerical simulations and experiments have provided insights on the mechanisms underlying these patterns and their relationships among each other \cite{Halatek.etal2018,Brauns.etal2020a}.

Here, we revisit the comparatively simple pole-to-pole oscillation employing the local-equilibria theory in the two-compartment setting.
This offers a fresh perspective on the Min-protein dynamics as it allows us to understand this elementary dynamic pattern in terms of phase space geometry, independently of numerical simulations.
In future work, this could serve as a starting point to systematically understand more complex patterns, like ``stripe oscillations'' (standing waves) in filamentous cells \cite{Raskin.deBoer1999,Halatek.Frey2012} and the zoo of patterns found \emph{in vitro} \cite{Vecchiarelli.etal2016,Glock.etal2019,Brauns.etal2020a}.

Intuitively, the two-compartment system represents the two cell poles (or cell halves) of the rod-shaped \textit{E.~coli} bacterium, as shown in Fig.~\ref{fig:motivation}a (see Appendix~\ref{app:3D-to-compartments} for a systematic reduction starting from the full three-dimensional cell geometry).
Figure~\ref{fig:motivation}c,d shows that the key qualitative features of Min pole-to-pole oscillations are still captured by the two-compartment model (see also Fig.~\ref{fig:Min-systematic-reduction} in Appendix~\ref{app:Min}).
While this two-compartment model cannot be expected to give a detailed quantitative description of Min oscillations, it has the advantage of informing about the basic underlying mechanism.
This complements earlier quantitative studies of the \textit{in vivo} dynamics \cite{Halatek.Frey2012,Wu.etal2016}.
Moreover, the two-compartment model serves as a minimal system for an oscillation mode recently reported for an \textit{in vitro} reconstitution of the Min system in microfluidic devices \cite{Brauns.etal2020a}. There, the oscillations go back and forth between two membrane surfaces through the bulk solution in-between them (see Fig.~\ref{fig:Min-in-vitro-reduction} in Appendix.~\ref{app:Min}). The analogy between this \textit{in vitro} oscillation mode and pole-to-pole oscillations \textit{in vivo} is further discussed in the conclusions, Sec.~\ref{sec:Min-conclusions}. 

We use a well-established minimal model for the Min-protein interactions that has been shown to successfully reproduce and predict a large range of experimental findings, quantitatively \emph{in vivo} and qualitatively \emph{in vitro} \cite{Huang.etal2003, Halatek.Frey2012, Halatek.etal2018, Denk.etal2018, Brauns.etal2020}.
For a detailed description of the model, we refer the reader to Refs.~\cite{Halatek.Frey2012,Brauns.etal2020}.
In short, the minimal model employs mass-action law kinetics to account for the attachment and detachment of MinD and MinE to and from the membrane and for their interactions there (see Fig.~\ref{fig:motivation}b): Membrane-bound MinD amplifies the attachment of further MinD from the cytosol with rate $\kdD$ and also recruits MinE from the cytosol with rate $\kdE$ to form MinDE complexes on the membrane. In these complexes, MinE stimulates MinD hydrolysis with rate $\kde$, leading to the dissociation of the complex and detachment of both proteins to the cytosol. 
In the cytosol, MinD undergoes nucleotide exchange from the ADP-bound form to the ATP-bound form, which can then attach to the membrane again.

Mathematically, the above reaction kinetics are described by the system of equations of the form Eq.~\eqref{eq:two-compartment-dyn} with
$\uvec = (\md, \mde, \cDT, \cDD, \cE)$, $\mathfrak{D} = \diag (\Dhd, \Dhde, \DhD, \allowbreak \DhD, \DhE)$ and 
\begin{equation} \label{eq:Min-reactions}
	\fvec(\uvec) = 
	\begingroup \renewcommand*{\arraystretch}{1.2} \begin{pmatrix}
	\phantom{-}R_\mathrm{D}^\mathrm{on} (\uvec) - R_\mathrm{E}^\mathrm{on} (\uvec) \\
	\phantom{-}R_\mathrm{E}^\mathrm{on} (\uvec) - R_\mathrm{DE}^\mathrm{off} (\uvec) \\
	-R_\mathrm{D}^\mathrm{on} (\uvec) + \lambda \cDD \\
	R_\mathrm{DE}^\mathrm{off} (\uvec) - \lambda \cDD \\
	-R_\mathrm{E}^\mathrm{on} (\uvec) + R_\mathrm{DE}^\mathrm{off} (\uvec)
	\end{pmatrix} \endgroup,
\end{equation}
where the reaction terms
\begin{subequations}
	\begin{align}
		R_\mathrm{D}^\mathrm{on} (\uvec) &= (\kD + \kdD\md)\cDT \, , \\
		R_\mathrm{E}^\mathrm{on}(\uvec) &= \kdE\md\cE \, , \\
		R_\mathrm{DE}^\mathrm{off}(\uvec) &= \kde\mde \, ,
	\end{align}
\end{subequations}
account, respectively, for MinD attachment and self-recruitment to the membrane, MinE recruitment by MinD, and dissociation of MinDE complexes with subsequent detachment of both proteins to the cytosol. 
The term $\lambda \cDD$ accounts for nucleotide exchange, i.e.\ conversion from $\cDD$ to $\cDT$, in the cytosol.
Importantly, these reaction kinetics conserve the total number of MinD and MinE proteins, $\bar{n}_\mathrm{D}^{}$ and $\bar{n}_\mathrm{E}^{}$, individually, i.e.\ there are \textit{two} globally conserved masses that are redistributed between the two compartments (cell halves) 
\footnote{In the most general form, each such conservation law can be written as $\mathbf{s}^\mathsf{T} {\cdot} \fvec = 0$, where the entries in the ``stoichiometric'' vector $\mathbf{s}$ account for the components whose sum is conserved. In the case of the Min system, we have $\mathbf{s}_\mathrm{D}^\mathsf{T} = (1,1,1,1,0)$ and $\mathbf{s}_\mathrm{E}^\mathsf{T} = (0,1,0,0,1)$ for the conservation of MinD and MinE respectively.}.

Numerically integrating the above set of ordinary differential equations using the parameters from Ref.~\cite{Halatek.Frey2012} yields pole-to-pole oscillations in good qualitative agreement with the oscillations found in the full three-dimensional geometry (see Fig.~\ref{fig:Min-systematic-reduction}a,b in Appendix~\ref{app:Min}).
Importantly, these oscillations persist if diffusive exchange between the compartments is set to a slow time scale compared to the reaction kinetics (see Fig.~\ref{fig:Min-systematic-reduction}c)
In this limit, the concentrations in the two compartments adiabatically follow the equilibrium concentrations that depend on the local masses
$n_{\mathrm{D},i}^{}, n_{\mathrm{E},i}^{}$
in the two compartments.
Hence, we can again apply the LQSSA, Eq.~\eqref{eq:LQSSA}, substituting the concentrations $\uvec$ by the reactive equilibria $\uvec^*$.
A discussion of the validity of this approximation and potential generalizations is deferred to the Conclusions, Sec.~\ref{sec:conclusions}. 

\begin{figure*}[tbp]
	\centering
	\hspace*{-1cm}\includegraphics{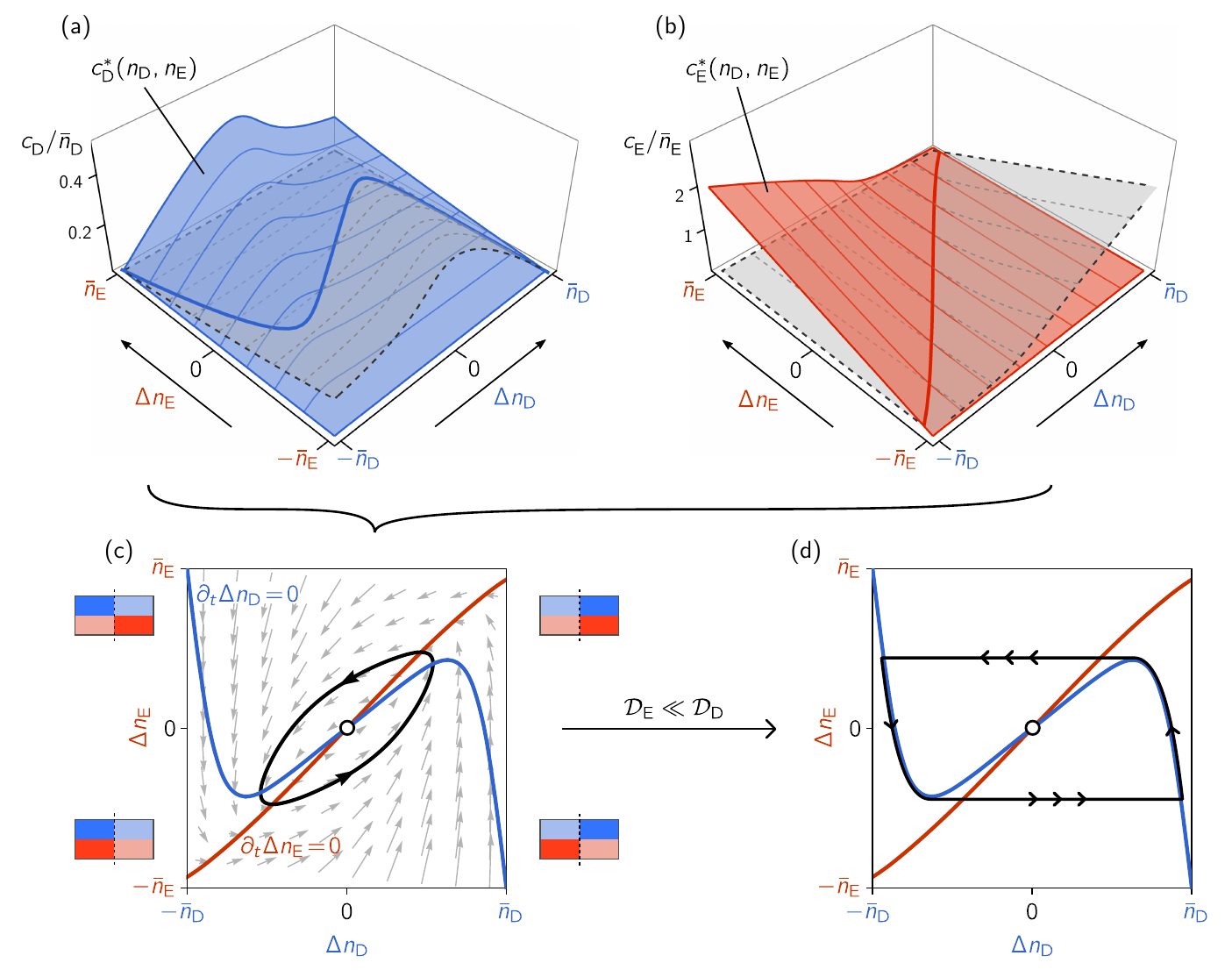}
	\caption{%
	Graphical construction of the dynamics in control space from the reactive nullcline surfaces.
	(a,b) Reactive nullcline surfaces showing MinD and MinE cytosol concentration (shaded blue and red respectively) as a function of the mass differences $\dnD,
	\dnE$. The intersection of each surface with its point reflection (shaded in gray with dashed outlines) determine the mass-redistribution nullclines (see text for details).
	These nullclines are a generalization of the fixed points shown in Fig.~\ref{fig:2cMcRD-NC-construction}b,c.
	(c) Phase portrait of the dynamics Eq.~\eqref{eq:Min-control-space-dyn-simple} with the MinD and MinE mass-redistribution nullclines obtained by the construction in (a) and (b) and the limit cycle trajectory (black) corresponding to pole-to-pole oscillations (cf.\ Fig.~\ref{fig:motivation}e).
	(d) Setting MinE diffusion to a slower time scale transforms the limit cycle trajectory to the shape characteristic for relaxations oscillations. 
	}
	\label{fig:Min-surface-construction}
\end{figure*}

The reactive equilibria as a function of the masses $\nD$ and $\nE$ are (for each compartment) determined by (cf.\ Eq.~\eqref{eq:loc-eq})
\begin{equation} 
\label{eq:Min-loc-eq}
	\mathbf{u}^*(\nD,\nE) : \left\{
	\begin{array}{l}
		\fvec\big(\mathbf{u}^*\big) = 0 \, ,\\
		\cD^* + \md^* + \mde^* = \nD \, ,\\
		\cE^* + \mde^* = \nE \, ,
	\end{array}
	\right.
\end{equation}
where we introduced the total cytosolic MinD concentration $\cD = \cDD + \cDT$.
For each component in the concentration vector $\uvec^*$ this defines a surface parametrized by $\nD$ and $\nE$, as shown in Figure~\ref{fig:Min-surface-construction}a,b for $\cD^*$ and $\cE^*$ (the respective surfaces for the membrane concentrations $\md^*$ and $\mde^*$ are shown in Fig.~\ref{fig:Min-md-mde-surfaces} in Appendix~\ref{app:Min}). 
We will term these \emph{reactive nullcline surfaces}.
In the following, we show how the dynamics of the local masses
$n_{\mathrm{D},i}^{}, n_{\mathrm{E},i}^{}$
can be inferred from these surfaces, analogously to the construction shown in Fig.~\ref{fig:2cMcRD-NC-construction} for two-component MCRD systems.

Because the total number of MinD and MinE proteins are conserved, only the protein masses redistributed between the two polar zones, $\dnDE(t)$, are time dependent and the mass densities of MinD and MinE in the right and left polar zone are given by 
%
\begin{equation} \label{eq:redistributed-mass}
	n_{\alpha,1/2} (t) = \bar{n}_\alpha \pm \Delta n_\alpha (t), \quad \alpha = \mathrm{D,E}.
\end{equation}
Analogously to the two-component system, we call the redistributed masses $\dnDE(t)$ the \emph{control variables} and the $(\dnD,\dnE)$-phase plane the \emph{control space}.
The dynamics in control space are governed by
\begin{subequations} \label{eq:Min-control-space-dyn}
\begin{align}
	\partial_t \dnD(t) 
		&= -\DhD \, \dcD - \Dhd \, \dmd - \Dhde\, \dmde, \label{eq:dnD-flow} \\
	\partial_t \dnE(t) 
		&= -\DhE \, \dcE - \Dhde\, \dmde, \label{eq:dnE-flow} 
\end{align}
\end{subequations}
where the concentration gradients (differences between the two polar zones) of the local equilibria are defined as (cf.\ Eq.~\eqref{eq:scaffolded-gradients})
\begin{equation}
	\Delta \uvec^*(\Delta n) := \uvec^*(\bar{n} + \Delta n) - \uvec^*(\bar{n} - \Delta n).
\end{equation}





\subsection{From reactive nullcline surfaces to mass-redistribution nullclines}

To understand the qualitative structure of the control-space dynamics Eq.~\eqref{eq:Min-control-space-dyn}, we first consider the lines along which there is no mass-redistribution of MinD/E respectively, $\partial_t \dnDE = 0$.
We term these \emph{mass-redistribution} nullclines.
Importantly, these are not to be confused with the \emph{reactive} nullcline (line of reactive equilibria) along which the reactive flow vanishes within a single compartment.

As we shall see in Sec.~\ref{sec:Min-relaxation-oscillation}, one can neglect the slow membrane diffusion to understand the basic oscillation mechanism of the Min system. We therefore consider this simpler case, $\Dhd = \Dhde = 0$, first. 
Equation~\eqref{eq:Min-control-space-dyn} then reduces to 
\begin{subequations} 
\label{eq:Min-control-space-dyn-simple}
\begin{align}
	\partial_t \dnD(t) 
		&=  -\DhD \, \dcD(\dnD,\dnE),	
    \label{eq:dnD-flow-simple} \\
	\partial_t \dnE(t) 
		&= -\DhE \, \dcE(\dnD,\dnE),
    \label{eq:dnE-flow-simple} 
\end{align}
\end{subequations}
describing how mass redistribution is driven by the gradients in the cytosolic protein densities, which are slaved to the local equilibria.
Thus, the mass-redistribution nullclines are simply given by $\Delta c^*_\mathrm{D} = 0$ and $\Delta c^*_\mathrm{E} = 0$. 
Geometrically, this corresponds to the intersection lines between the reactive nullcline surfaces $c_\mathrm{D,E}^*(\nD,\nE)$, and their respective point reflections, reflected at the point $(\bar{n}_\mathrm{D}, \bar{n}_\mathrm{E})$; see gray surfaces with dashed outlines in Fig.~\ref{fig:Min-surface-construction}a,b. 
In other words, the shape of reactive nullcline surfaces encodes the essential information about the nonlinear reaction kinetics for the dynamics of the spatially coupled system.

This construction is the analog to the construction for the two-component system shown in Fig.~\ref{fig:2cMcRD-NC-construction}.
In fact, in slices with $\nE = \mathrm{const}$, the line $\cD^*(\nD)$ has the same shape as the nullcline shown in Fig.~\ref{fig:2cMcRD-NC-construction}.
This ``hump shape'' gives rise to the N-shaped MinD-redistribution nullcline ($\partial_t \dnD = 0$, see blue line in Fig.~Fig.~\ref{fig:Min-surface-construction}a,c). 
The two outer branches of this N-shaped nullcline represent polarized MinD states, corresponding to the two outer fixed points in the analogous two-component system construction Fig.~\ref{fig:2cMcRD-NC-construction}c.
We will make this more concrete below in Sec.~\ref{sec:MinE-decoupled}.
If $\bar{n}_\mathrm{D}$ lies to the left of the crest of $\cD^*(\nD,\nE)$, the resulting MinD-redistribution nullcline is monotonic, analogous to the single fixed point in Fig.~\ref{fig:2cMcRD-NC-construction}b. 
The crest of the $\cD^*$ surface defined by $\partial_{n_\mathrm{D}} \cD^*(n_\mathrm{D},n_\mathrm{E}) = 0$ approximately follows the line $\nE/\nD \approx \kdD/\kdE$ for sufficiently large $\nE$ (specifically in the limit $\nE^2 \gg \kD \kdD \kde/\kdE^3$). This relation is found by applying the implicit function theorem to evaluate the derivative $\partial_{n_\mathrm{D}} \cD^*$ using the definition Eq.~\eqref{eq:Min-loc-eq} for the reactive equilibria.

In contrast to the non-trivial MinD-redistribution nullcline, the monotonicity of the surface $\cE^*(\nD,\nE)$ gives rise to a monotonic MinE-redistribution nullcline (red line in Fig.~\ref{fig:Min-surface-construction}b,c) for all $\bar{n}_\mathrm{D},\bar{n}_\mathrm{E}$.

\textit{Mass-redistribution potentials.\,}
In passing, let us introduce an alternative formulation of the mass-redistribution dynamics Eq.~\eqref{eq:Min-control-space-dyn} that allows one to generalize the graphical construction presented in Fig.~\ref{fig:Min-surface-construction} to arbitrary values of all diffusion constants (including $\mathcal{D}_\mathrm{d,de} > 0$).
Using the mass-redistribution potentials (cf.\ Ref.~\cite{Brauns.etal2020b}), $\eta_\mathrm{D} = \cD + (\Dhd/\DhD) \md + (\Dhde/\DhD) \mde$ and $\eta_\mathrm{E} = \cE + (\Dhde/\DhE) \mde$, Eq.~\eqref{eq:Min-control-space-dyn} can be written as
\begin{subequations}
\begin{align}
	\partial_t \dnD(t) 
		&= -\DhD \, \Delta \eta^*_\mathrm{D}, \\
	\partial_t \dnE(t) 
		&= -\DhE \, \Delta \eta^*_\mathrm{E}. 
\end{align}
\end{subequations}
Since these equations have the same form as Eq.~\eqref{eq:Min-control-space-dyn-simple}, the construction shown in Fig.~\ref{fig:Min-surface-construction} can be generalized by exchanging $c_\mathrm{D,E}^*$ for $\eta_\mathrm{D,E}^*$. 
The surfaces $\eta^*_\mathrm{D}$ and $\eta^*_\mathrm{E}$ can be interpreted as ``superpositions'' of the local-equilibrium surfaces of the individual components weighted by the respective exchange rates $\mathcal{D}_i$.
The effect of reaction rates or diffusion constants on the spatial dynamics is encoded in the deformation of these surfaces under variation of these parameters (see Movies~1 and~2).

\subsection{Min pole-to-pole oscillations are relaxation oscillations}
\label{sec:Min-relaxation-oscillation}

The nullclines enable one to read off the qualitative structure of the dynamics in the $(\dnD,\dnE)$-phase plane \cite{StrogatzBook,JacksonBook}.
Specifically, one immediately recognizes the characteristic scenario of a relaxation oscillator
\footnote{Relaxation oscillators are often encountered in simple two-component models of (biological) oscillators and switches; see e.g.\ the FitzHugh--Nagumo model \cite{FitzHugh1961,Nagumo.etal1962} and Chapter 5 in Ref.~\cite{Murray2002}.}.
Recalling that the two outer branches of the N-shaped MinD-redistribution nullcline correspond to polarized MinD states, this shows that Min pole-to-pole oscillations are \emph{relaxation oscillations of the MinD-polarity direction} driven by mass-redistribution of MinE between the two cell halves.

The limit cycle of relaxation oscillators can be graphically constructed in the limit where the variable with the N-shaped nullcline evolves on a much faster time scale compared to the other variable \cite{JacksonBook}.
In the Min system, this corresponds to setting MinE redistribution to a much slower time scale than MinD redistribution ($\DhD \gg \DhE$).
In this limit, the limit cycle deforms into a ``trapezoidal'' trajectory; see Fig.~\ref{fig:Min-surface-construction}d. The dynamics slowly follows the N-shaped MinD-redistribution nullcline (polarized MinD states), driven by MinE redistribution, and rapidly switches between the left and right branches at the extrema of this nullcline, driven by MinD redistribution.

In a broader context, the above analysis demonstrates how the reactive nullcline surfaces and their intersection lines---which are the mass-redistribution nullclines---are helpful tools to explore the ability of systems to show nontrivial spatial dynamics without the need to perform a full scale finite element simulation.
The shape of the reactive nullcline surfaces and thus the mass-redistribution nullclines are ultimately a consequence of the nonlinear feedback in the reaction kinetics.
In the specific case of the Min system, these are the recruitment terms $\kdD \md \cD$ and $\kdE \md \cD$.
It is important to recall that the shape of the nullclines resulting from the reaction kinetics, and not the specific reaction kinetics per se, determines the spatial (mass-redistribution) dynamics.
Hence, different reaction terms can give rise to same nullcline geometry, and thus the same spatial dynamics.
Rather than classifying dynamics based on their reaction network topology, this suggests that a classification might be possible in terms of the shapes of their reactive nullcline surfaces and the resulting mass-redistribution nullclines. 
We demonstrate this in Sec.~\ref{sec:PAR-and-Cdc42}, where we analyze two further paradigmatic models for intracellular pattern formation.

\subsection{The role of diffusive MinE transport}
\label{sec:MinE-decoupled}

So far, we have neglected membrane diffusion to elucidate the basic Min-oscillation mechanism. 
We now relax that approximation and first consider the role of MinE membrane diffusion. 
Using the conservation law $\mde + \cE = \nE$, Eq.~\eqref{eq:dnE-flow} can be recast as
\begin{equation} \label{eq:MinE-redistri}
	\partial_t \dnE = -(\DhE - \Dhde) \Delta \cE^*(\dnD,\dnE) - \Dhde \dnE.
\end{equation}
This shows that diffusive transport on the membrane always counteracts cytosolic transport. 
In particular, if one were to set $\DhE = \Dhde$, there would be no MinE mass-redistribution since Eq.~\ref{eq:MinE-redistri} would reduce to $\partial_t \dnE = -\Dhde \dnE$, such that $\dnE$ would simply relax to the homogeneous state $\dnE = 0$.
Thus, in control space, the MinE-redistribution nullcline would simply be given by $\dnE = 0$, which intersects the N-shaped MinD nullcline at three points, representing the unstable homogeneous steady state in the center and two stable polarized states on the left and right, respectively.
Hence, the dynamics would reduce to the one-dimensional control space for MinD redistribution which, corresponding to the scenario shown in Fig.~\ref{fig:2cMcRD-NC-construction}.
From this Gedankenexperiment, we conclude that the elementary pattern-forming mechanism of the Min system is MinD polarization and does not require spatial redistribution of MinE.
The specific function of MinE in MinD polarization is that of a ``local catalyst'' that provides nonlinear feedback essential in shaping the non-monotonic reactive MinD nullcline $\cD^*(\nD)$.
Thus, while redistribution of MinE is not required for the formation of a polarized MinD pattern, it causes the emergence of oscillations by periodically inducing switching of the MinD polarization direction as we showed in the previous section.

Physiologically, $\DhE = \Dhde$ would actually correspond to a scenario where free MinE remains membrane bound, i.e.\ MinE would cycle between the MinD-bound and the free conformation on the membrane and $\cE$ would then denote the concentration of free MinE. 
The stationary patterns resulting in this case provide a potential hint for the possible biomolecular features of MinE responsible for the (quasi-)stationary patterns reported in recent experiments using MinE purified with a His-tag at the C-terminus instead of the N-terminus \cite{Glock.etal2019}.
Compared to his-MinE, MinE-his might have a strong membrane affinity causing free MinE to remain membrane-bound after the dissociation of MinDE complexes. Free MinE on the membrane diffuses much slower than in the cytosol thus suppresses the MinE redistribution that gives rise to dynamic patterns (waves and oscillations).
This hypothesis suggests that increasing the MTS strength of MinE might cause a transition from dynamic to quasi-stationary patterns.

\begin{figure*}[tbp]
	\centering
    \includegraphics{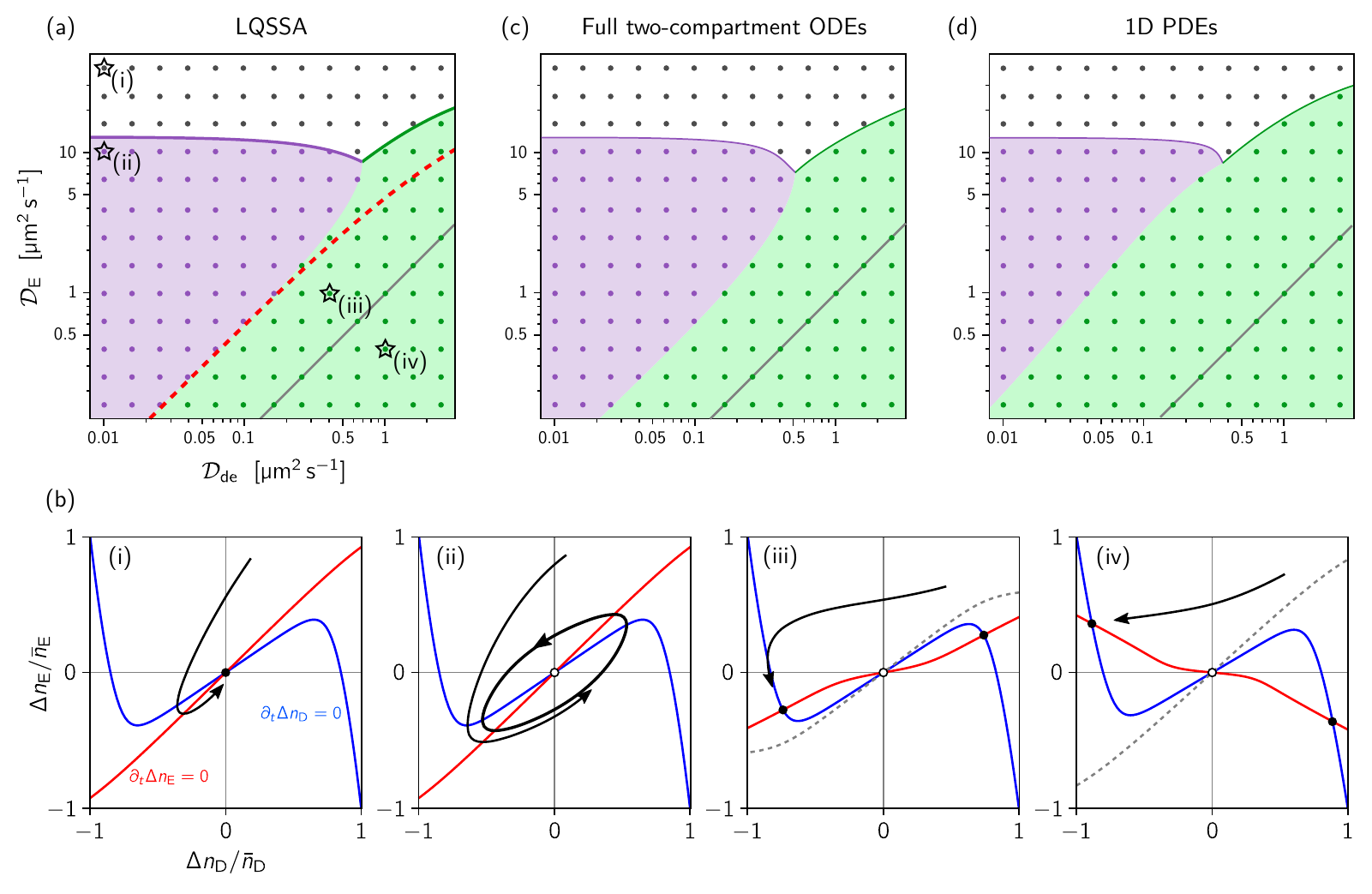}
	\caption{%
	Phase diagrams of Min protein dynamics. In each panel, the points and shaded background indicate the results from numerical simulations, distinguishing no patterns (gray), oscillations (purple) and stationary polarity (green). 
	(a) Phase diagram for the LQSSA dynamics Eq.~\eqref{eq:Min-control-space-dyn}. 
	The solid purple and green line indicate the Hopf and pitchfork bifurcations found by linear stability analysis of the LQSSA dynamics. Along the dashed red line, the MinE-redistribution nullcline intersects MinD-redistribution nullcline at the latter's extrema. In the limit $\DhE,\Dhde \ll \DhD$, this marks the transition between relaxation oscillations and stationary polarity. The gray line indicates the line $\DhE = \Dhde$. 
	(b) Example trajectories in the $(\dnD,\dnE)$-phase plane (cf.\ Fig.~\ref{fig:Min-surface-construction}c): (i) no instability, (ii) pole-to-pole oscillations, (iii) stationary polarity for $\DhE > \Dhde$, (iv) stationary polarity for $\DhE < \Dhde$ (note the opposite slope of the MinE-redistribution nullcline). In (iii) and (iv), the dashed gray line shows the separatrix, that separates the basins of attraction of the two polarized states.
	(c) Phase diagram from numerical simulations of the full two-compartment dynamics, Eq.~\eqref{eq:two-compartment-dyn}. Note the excellent agreement with LQSSA (a).
	(d) Phase diagram from numerical simulations of the PDEs on a line (varying $D_\mathrm{E}$ in full three-dimensional cell geometry affects also the vertical gradients rather than just lateral diffusion).
	}
	\label{fig:Min-phase-diagram}
\end{figure*}

To elucidate the role of MinE transport more quantitatively, we now study the transition from stationary to oscillatory patterns as a function of the diffusion constants $\DhE$ and $\Dhde$.
Varying these diffusion constants results in a deformation of the MinE-redistribution nullcline in control space.
Specifically, the shape of the MinD-redistribution nullcline only depends on the difference $\DhE - \Dhde$, i.e.\ the balance of co-polarizing diffusion of free MinE compared to the contra-polarizing diffusion of MinDE complexes.
In the relaxation-oscillation limit where MinE-redistribution is much slower than MinD redistribution ($\Dhde,\DhE \ll \DhD$), the locations of the intersection points between the MinD's and MinE's mass-redistribution nullclines determine whether the system is oscillator or exhibits stationary polarity (see Fig.~\ref{fig:Min-phase-diagram}b). The transition case separating these two regimes is when the MinE-redistribution nullcline intersects the MinD-redistribution nullcline at its extrema.
In addition, the stability of the homogeneous steady state can be obtained by a linear stability analysis in LQSSA (see Appendix~\ref{app:LSA}). The resulting ``phase diagram'' is shown in Fig.~\ref{fig:Min-phase-diagram}a.

This phase diagram obtained using LQSSA can be compared to the phase diagram of the full model obtained by numerical simulations (see Fig.~\ref{fig:Min-phase-diagram}b).
The fact that the topology of the two phase diagrams agrees shows that the reduced dynamics, Eq.~\eqref{eq:Min-control-space-dyn}, accounts for the relevant physics of the \emph{in vivo} Min system.

\subsection{Concluding remarks on the Min system.}
\label{sec:Min-conclusions}

We have shown that dynamics underlying Min pole-to-pole oscillations can be reduced to the redistribution of MinD and MinE mass between the two cell poles. 
A simple geometrical construction yields the qualitative phase space structure of the mass-redistribution dynamics. 
Specifically, we recovered the paradigmatic N-shaped nullcline that underlies general relaxation oscillations.
This systematic reduction immediately allowed us to transfer the knowledge on relaxation oscillations to the Min pole-to-pole oscillations.
The outer legs of the N-shaped MinD-redistribution nullcline correspond to oppositely polarized MinD states.
MinE redistribution drives cyclic switching between these two states, giving rise to the pole-to-pole oscillations.
In the absence of MinE redistribution (achieved by setting $\DhE = \Dhde$), MinD forms stationary polarized patterns instead.
This shows that the elementary pattern underlying for pole-to-pole oscillations in the \emph{in vivo} Min system is not oscillatory but generic cell polarity. 
We conclude that the oscillatory dynamics are not a direct property of the kinetic interaction network, which is the same for the oscillatory and non-oscillatory regime.
Instead, oscillations arise as consequence of MinE redistribution ``downstream'' of MinD polarization. MinE redistribution is not necessary for MinD polarization while MinD redistribution is strictly required.
This links pole-to-pole oscillation in the Min system and generic cell polarity and suggests a hierarchy of species in large multi-species multi-component systems.
Notably, this also shows that the functional role of MinE for pattern formation cannot be considered to be that of an inhibitor in the sense of the ``activator--inhibitor'' mechanism \cite{Gierer.Meinhardt1972,Segel.Jackson1972}.

The above analysis of the mass-redistribution dynamics elucidates the different roles of MinD and MinE redistribution for Min-protein pattern formation.
In Sec.~\ref{sec:PAR-and-Cdc42}, we apply the same reduction approach to two other intracellular systems. 
This will allow us to compare the underlying pattern-forming mechanisms on the level of their mass-redistribution nullcline geometries.

\paragraph{Min oscillations \textit{in vitro}.} 
Let us emphasize again that the pole-to-pole oscillations emerge due to the diffusive coupling of two compartments, representing the two cell halves. An isolated compartment exhibits only stable, stationary states. In other words, the \emph{in vivo} Min system is not an ``oscillatory medium'' of coupled oscillators.
Remarkably, this is in stark contrast to the Min-protein pattern dynamics observed in classical \textit{in vitro} setups with a large cytosolic bulk volume on top of a flat membrane surface. Here, a single (laterally isolated) membrane patch is coupled to an extended cytosolic reservoir, and it is this coupling that gives rise to \emph{local} oscillations \cite{Halatek.etal2018,Brauns.etal2020a}. This shows that on a mechanistic level, Min protein patterns in cells are distinct from patterns in reconstituted systems with a large bulk. 

In a recent work, the Min system was studied in microfluidic chambers with two flat membrane surfaces separated by a bulk solution \cite{Brauns.etal2020a} (see Fig.~\ref{fig:Min-in-vitro-reduction} in Appendix~\ref{app:Min}).
This limits the bulk volume above each membrane patch and thus suppresses the local oscillations for sufficiently low bulk height. Interestingly, for intermediate bulk height,
experiments and a theoretical analysis have revealed an oscillation mode that transports mass between the two opposite membrane surfaces through the bulk in-between them. This oscillation is analogous to the \textit{in vivo} pole-to-pole oscillation where the two opposite membrane patches play the role of the cell poles \textit{in vivo}.
Correspondingly, with regard to the \emph{in vitro} geometry, the two-compartment system serves as a minimal system to represent single vertical bulk column and the membrane patches at its top and bottom; see Fig.~\ref{fig:Min-in-vitro-reduction}.

\paragraph{Historic note: Oscillations driven by diffusive coupling of two ``dead'' cells.}

Intriguingly, the Min-oscillation mechanism described above has some parallels to a conceptual model for diffusion-driven oscillations studied by Smale already in 1974 \cite{Smale1974}.
Smale's motivation, inspired by Turing's pioneering work \cite{Turing1952}, was to show how two identical reactors that exhibit only a stable stationary state when isolated, start oscillating (in anti-phase) when coupled diffusively. 
Or, as Smale put it: ``One has two dead (mathematically dead) cells interacting by a diffusion process which has a tendency in itself to equalize the concentrations. Yet in interaction, a state continues to pulse indefinitely.''
As we showed above, the \emph{in vivo} Min system also has that property.

Remarkably, Smale used a relaxation oscillator as starting point to construct the diffusion driven two-compartment oscillator.
In a broader view, this demonstrates how structures in phase space, like fixed points and nullclines, are powerful tools to understand and design nonlinear systems. For instance, they have been used to great success in the study of neuronal dynamics \cite{Izhikevich2007} and biochemical oscillators \cite{Novak.Tyson2008,Ferrell.etal2011}.

\section{Control space flow of the PAR and Cdc42 systems}
\label{sec:PAR-and-Cdc42}

\begin{figure*}[tbp]
	\centering
	\includegraphics{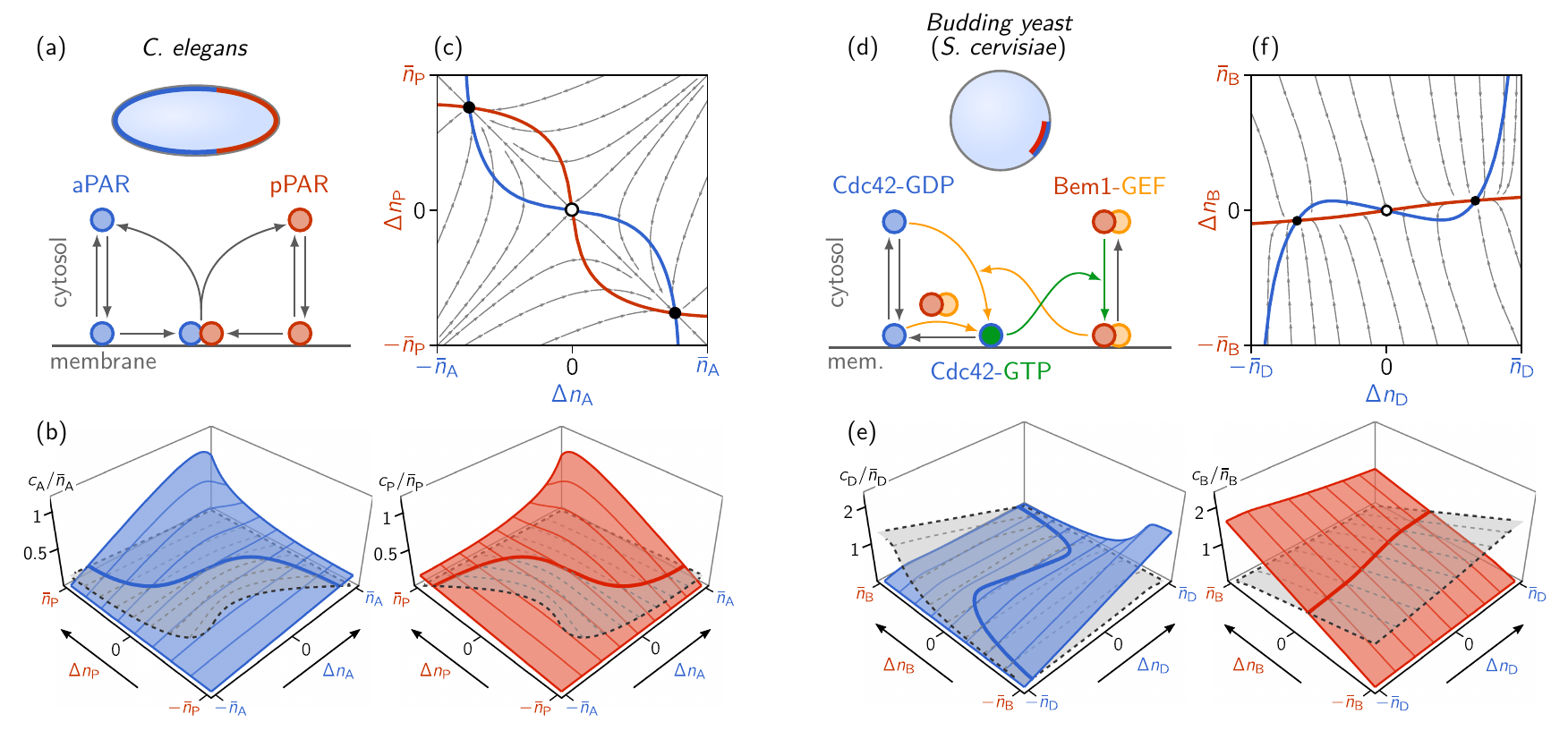}
	\caption{%
	Reaction networks, reactive nullcline surfaces and control-space phase portraits for the PAR system of \textit{C.\ elegans} (a--c) and the Cdc42 system of \textit{S.\ Cerevisiae} (d--f).
	(a) Cartoon of a \textit{C.\ elegans} embryo showing the segregated aPAR and pPAR domains which as a result of mutual detachment of aPAR and pPAR proteins from the membrane.
	(b) Reactive nullcline surfaces of aPARs (blue, left) and pPARs (red, right). Note the symmetry under the exchange A$\leftrightarrow$P.
	(c) Control-space phase portrait showing the mass-redistribution dynamics and the mass-redistribution nullclines of aPARs (blue) and pPARs (red). Both mass-redistribution nullclines intersect the lines $\Delta n_\mathrm{A,P} = 0$ only once, indicating that pattern formation requires redistribution of both protein species.
	(d) Cartoon of a budding yeast cell showing a polar cluster of co-localized Cdc42 and Bem1-GEF complexes. In WT cells, Cdc42 and Bem1-GEF complexes (homologous to Scd1-Scd2 complexes in fission yeast) mutually recruit each other to the membrane and are therefore co-localized in the resulting pattern.
	(e) Reactive nullcline surfaces of Cdc42 (blue, left) and Bem1-GEF (red, right).
	(f) Control-space phase portrait showing the mass-redistribution dynamics and the mass-redistribution nullclines of Cdc42 (blue) and Bem1-GEF (red). The N-shaped Cdc42-redistribution nullcline intersects the line $\Delta n_\mathrm{B} = 0$ three times, indicating that redistribution of Bem1-GEF complexes is not required for pattern formation. In contrast to MinE in the Min system, where cytosolic MinE redistribution drives oscillations, the cytosolic redistribution of Bem1-GEF complexes has a stabilizing effect on stationary patterns.
	}
	\label{fig:PAR-and-Cdc42}
\end{figure*}

The above investigation of the Min system demonstrates that the key characteristics of the spatio-temporal protein dynamics, and the underlying pattern-forming mechanisms, can be inferred from the shapes of the reactive nullcline surfaces.
In the following, we use the approach introduced above to two paradigmatic model systems for intracellular self-organization: the PAR system of \textit{C.\ elegans} and the Cdc42 system of budding yeast (\textit{S.\ cerevisiae}).
Starting from previously established mathematical models on spatially continuous domains, we follow the same reduction procedure as for the Min system; details on the models, parameter choices and the reduction procedure are described in Appendix~\ref{app:PAR-and-Cdc42}. Put briefly, the spatially continuous dynamics is mapped to the two-compartment setting, and the LQSSA is applied such that only the redistributed masses remain as dynamic variables.
The mass-redistribution dynamics can then be analyzed in terms of the reactive nullcline surfaces and the resulting phase portraits as shown in (Fig.~\ref{fig:PAR-and-Cdc42}).
This allows us to compare the pattern forming mechanisms underlying these different systems.

\paragraph{PAR system.}
The first division of \textit{C.\ elengans} embryos is asymmetric, where the fate of the daughter cells is defined by proteins called aPARs and pPARs that segregate along the long axis of the ellipsoidal cells \cite{Lang.Munro2017}.
A model for the formation of these segregated domains was introduce in Ref.~\cite{Goehring.etal2011}, based on the mutual antagonism between cortex-bound A- and pPARs (see Fig.~\ref{fig:PAR-and-Cdc42}a). Here, we adopt this model here, to illustrate the phase portrait structure that is characteristic of the mutual-antagonism mechanism. Model details and the parameters are given in Appendix~\ref{app:PAR-and-Cdc42}. 
Since, the reaction network (and the parameters used) are symmetric, so are the reactive nullcline surfaces (Fig.~\ref{fig:PAR-and-Cdc42}b).
From the resulting mass-redistribution nullclines (Fig.~\ref{fig:PAR-and-Cdc42}c), we can immediately see that the patterns form by segregation into domains where pPAR concentration is high while aPAR concentration is low and vice versa.
Notably, the mass-redistribution nullclines do not intersect the lines $\Delta n_\mathrm{A} = 0$ and $\Delta n_\mathrm{P} = 0$ away from the origin, indicating that PAR-pattern formation the requires the redistribution of both protein species.
Moreover, the topology of the phase portrait is such that oscillations cannot occur. 
We expect that these qualitative insights generalize to  more detailed models for PAR-protein polarity, see e.g.\ Ref.~\cite{Gessele.etal2020}.

\paragraph{Cdc42 system.}
Budding yeast cells divide asymmetrically by budding and growing a daughter cell. The division site is determined by the polarization of GTP-bound Cdc42 to a ``spot'' on the membrane \cite{Bi.Park2012}. 
In wild-type cells, Cdc42 polarization is driven by a mutual-recruitment mechanism that is facilitated by the scaffold protein Bem1. Bem1 is recruited to the membrane by Cdc42-GTP. Membrane-bound Bem1 then recruits Cdc42's GEF, Cdc24, forming Bem1-GEF complexes. In turn, Bem1-GEF complexes recruit Cdc42-GDP from the cytosol and catalyze its conversion to Cdc42-GTP, thus closing the feedback loop.
To illustrate the phase portrait structure that is characteristic of this mutual recruitment mechanism, we adopt a simplified form of the detailed, quantitative model introduced in Ref.~\cite{Klunder.etal2013}; see Appendix~\ref{app:PAR-and-Cdc42}.
In the simplified model, Bem1-GEF complexes are described as a single species with a membrane-bound and a cytosolic state (see Fig.~\ref{fig:PAR-and-Cdc42}d). 
Figure~\ref{fig:PAR-and-Cdc42}e,d shows the reactive nullcline surfaces and the resulting phase portrait of this model.
The location of the mass-redistribution nullcline intersection points, corresponding to stationary polarized states, indicates that Cdc42 and Bem1-GEF complexes co-polarize as expected.
Moreover, the N-shaped Cdc42-redistribution nullcline that intersects the line $\Delta n_\mathrm{B} = 0$ three times, indicating that polarization does not require spatial redistribution of Bem1-GEF complexes. Still, the enzymatic action of Bem1-GEF complexes in the local reaction kinetics is essential for Cdc42 polarization as they provide the nonlinear feedback that shapes the Cdc42-redistribution nullcline.
In this sense, Bem1-GEF complexes play an analogous but inverse role to MinE in the Min system. In the physiological case, Bem1-GEF complexes they stabilize polarity by co-polarizing with Cdc42. In the unphysiological case that free Bem1-GEF complexes diffuse slower that membrane-bound ones ($\mathcal{D}_\mathrm{b} > \mathcal{D}_\mathrm{B}$), contra-polarization of Bem1-GEF complexes drives cycling switching of Cdc42 polarity. 
Thus, the Cdc42 system and the Min system can be regarded as two complementary versions of the same mechanism in which the enzymatic function of the ``secondary protein'' (Bem1-GEF/MinE) is reversed such that its spatial redistribution has opposite effects in the two systems.

The above analysis has a striking implication: On the level of the pattern forming mechanisms, the Cdc42 system is closely related to the Min system, while the PAR system operates based on a fundamentally different mechanism. From the perspective of the phenomenology exhibited for physiological parameters, this is highly surprising since the Cdc42 system and the PAR system exhibit stationary polarity patterns, while the Min system exhibits pole-to-pole oscillations.

\section{Conclusions}
\label{sec:conclusions}

Quantitative models of biological systems are typically multi-component multi-species systems with a high-dimensional parameter space. It is therefore particularly challenging to find a unifying level of description where the mechanisms underlying different models can be compared.

Here, we presented a reduction method to obtain a phase-portrait representation of mass-conserving pattern forming systems which crystallizes their key qualitative features. 
This reduction is based on two steps.
First, a reduction of the spatially continuous domain to two well-mixed compartments coupled by diffusion. This approximation assumes that the pattern of interest is a single ``interface'' connecting a high density region to a low density region. This is rather the rule than the exception for protein patterns observed in cells, especially bacterial cells due to their small size. Moreover, such an interface can also be interpreted as the elementary building block of more complex patterns with many interfaces. 
Second, the local quasi-steady-state approximation (LQSSA), which assumes that the relaxation of the concentrations in the compartments to a reactive equilibrium (local quasi-steady state) is fast compared to slow diffusive mass exchange between the compartments.
This approximation is motivated by the insights that the essential degree of freedom is the spatial redistribution of the conserved masses and that the key information about the reaction kinetics is encoded in the dependence of the reactive equilibria on these masses. Limitations and potential extensions of the LQSSA are discussed in the Outlook, Sec.~\ref{sec:outlook}.

After these two reduction steps, the only remaining degrees of freedom are the differences in globally conserved masses between the two compartments.
In this reduced system, the dynamics of these mass differences can simply be inferred from the reactive nullcline (hyper-)surfaces. Specifically, the intersection lines of \emph{reactive} nullcline surfaces act as \emph{mass-redistribution} nullclines in the phase space of the redistributed masses.
The mass-redistribution nullclines depend on the diffusion constants and thus inform about the role of mass-redistribution in the observed phenomena.
Thus, they allow a classification of pattern-forming systems, as we demonstrated by comparing the phase portraits of three different protein-pattern forming systems.
Attempts to classify pattern-forming systems based on the topology of the protein interaction network face the difficulty that many networks can give rise to similar phenomena, and the same network can produce different phenomena depending on parameters (e.g.\ stationary and oscillatory patterns in the Min system).
In contrast, here we have demonstrated that the geometry of the reactive nullcline surfaces informs on the key qualitative features of the observed dynamics.
This suggests that one can identify geometric design principles based on the shape of the reactive nullcline surfaces and the resulting mass-redistribution nullclines.
Such design principles might guide future model building efforts in a similar way as the design principles that have been identified for neural excitability \cite{Izhikevich2007} and well-mixed biochemical oscillators \cite{Novak.Tyson2008,Ferrell.etal2011}.

The phase-portrait analysis in terms of mass-redistribution nullclines also shows that not all species need to be redistributed for patterns to form in the first place. One can construct a ``core'' pattern-forming system, where these species are considered non-diffusible and their kinetics absorbed into effective kinetics of the redistributed species. In the Min system and the Cdc42 system, the (local) \emph{enzymatic action} of MinE / Bem1-GEF complexes is part of the core pattern-forming mechanism, whereas their \emph{cytosolic redistribution} is not. Redistribution of MinD / Cdc42 is sufficient for the formation of (stationary) MinD / Cdc42 patterns.
Thus, the elementary polarization mechanism is equivalent in the Min system and the Cdc42 system. The difference of these system lies in the effect of the mass redistribution of the ``secondary proteins'' MinE and Bem1-GEF respectively. 
In the Min system, redistribution of MinE by cytosolic diffusion system drives cyclic switching of the MinD polarity axis and thus gives rise to pole-to-pole oscillations. In contrast, redistribution of Bem1-GEF complexes stabilizes stationary Cdc42 polarization.

Taken together, the shape of the reactive nullcline surfaces and the resulting mass-redistribution nullclines inform about important qualitative features of a model and thus bridge the gap between nonlinear reaction kinetics and the observed phenomena. In particular, they allow one to disentangle the functional roles of each protein species in the pattern-forming mechanisms.

\paragraph{Assuming a well-mixed cytosol misses important physics.}

The assumption of a well-mixed cytosol is often made a priori, justified by the observation that diffusive transport on cellular scales is fast compared to membrane diffusion (and reaction kinetics); see e.g.\ \cite{Mori.etal2008, Holmes.etal2015, Diegmiller.etal2018, Xu.Jilkine2018, Guzzo.etal2018, Tostevin.etal2021}. 
This reasoning overlooks that the relative rates of transport can be important if there is more than one protein species diffusing in the cytosol. Or put differently, setting the cytosol concentrations well-mixed neglects that the cytosol gradients of different species can have different amplitudes, which may be mechanistically relevant, even if the cytosol gradients are shallow compared to membrane gradients.

In fact, for the Min system, we find that increasing the diffusion of free MinE eventually always suppresses pattern formation in the Min system (see phase diagram Fig.~\ref{fig:Min-phase-diagram}a and Appendix~\ref{app:Min-well-mixed-cytosol}).
This shows that the relative rate of cytosolic transport of MinD vs MinE (and, the relative amplitude of the cytosolic gradients, respectively) is important for the dynamics.
This shows that one misses important physics if one assumes a well-mixed cytosol a priori.

In general, the time scales of cytosol diffusion---even if fast--- and, correspondingly, the relative amplitudes of cytosolic gradients---even if shallow---can be important if there is more than one cytosolic (fast-diffusing) species. 
Approaches, such as the so called ``local-perturbation analysis'' \cite{Holmes.etal2015}, that rely on the a priori approximation to treat fast diffusing components as well-mixed,  may therefore miss important features of the dynamics.

In passing, we note that explicit cytosol diffusion is also important to account for effects due to cell geometry. This is relevant for the axis selection of polarity patterns in rod-shaped or ellipsoidal cells \cite{Thalmeier.etal2016, Glock.etal2019a, Gessele.etal2020}.
Compartment-based models---although requiring more than two compartments---have also been employed successfully to study such geometry effects~\cite{Thalmeier.etal2016}.

\subsection{Outlook}
\label{sec:outlook}

\paragraph{Future applications and generalizations.}

Going forward, it will be interesting to apply the reduction method and phase-portrait analysis presented here to other model systems, e.g.\ the oscillatory Cdc42-polarization in fission yeast \cite{Das.etal2012,Xu.Jilkine2018}.
The phase portrait analysis might be particularly helpful to study genuinely nonlinear phenomena like stimulus induced pattern formation and stimulus-induced polarity switching \cite{Tostevin.etal2021} which are not accessible to linear stability analysis.

Potential obvious generalizations of the two-compartment setting are systems with asymmetric exchange rates, and those with heterogeneous compartments (reaction kinetics, bulk-surface ratio, size).
Indeed, setting the redistribution of one species to a slow time scale in the models with two conserved masses (e.g.\ Min system), makes the system heterogeneous from the point of view of the fast species. The heterogeneity is determined by distribution of the slow species between the two compartments and changes on the slow time scale.
Concrete application for heterogeneous two-compartment models might be Ran-GTPase driven nuclear transport, where the two compartments represent the cytoplasm and nucleoplasm, with transport between them through pores in nuclear envelope \cite{Smith.etal2002,Gorlich.etal2003,Kim.Elbaum2013,Lolodi.etal2016}.
More broadly, two-compartment systems with asymmetric exchange rates and heterogeneous compartments have been studied in ecology \cite{Salomon.etal2010,Holt1985}, where interesting new effects compared to the symmetric case were found.

Another route of generalization is to study more than two coupled compartments. In this case, the phase space of the mass differences becomes high-dimensional and thus impractical for a phase-portrait analysis \cite{Thalmeier.etal2016}.
Instead, one can plot all local masses into one graph, as was done in Ref.~\cite{Halatek.etal2018}. This way, the spatial information is lost, but one can still gain insight about the role of the control space structure (surface of local equilibria and their stability) for the dynamics of the spatially coupled system.

\paragraph{Relation to parameter-space topology.}

A previous work on reaction--diffusion models for cell polarity has identified generic topological features of their parameter spaces \cite{Trong.etal2014}. 
In the specific case of two-component systems, the origin of these features was recently traced back to the phase space geometry, specifically the shape of the reactive nullcline of pattern forming systems (see Sec.~VII in Ref.~\cite{Brauns.etal2020b}). 
Two-compartment systems are a promising setting to generalize these findings to systems with more components and phenomena like pole-to-pole oscillations. Indeed, the way the mass-redistribution nullclines deform due to the variation of parameters (kinetic rates, diffusion constants, average masses) determines the bifurcations in parameter space. Thus, we expect a close relation between the geometry of mass-redistribution nullclines and phase space topology.

\paragraph{Relaxing the local quasi-steady state assumption.}

The analysis presented here relied crucially on the stability of the local equilibria and a time scale separation between reactive relaxation to the local equilibria and diffusive mass redistribution. This allowed us to make the LQSSA Eq.~\eqref{eq:LQSSA}.
In the absence of this time scale separation, the concentrations will deviate from the local equilibria due to the diffusive flows in the individual components.
For two-component systems, this deviation from the local equilibria has only a quantitative effect but does not change the dynamics qualitatively. This is because the local phase spaces are one-dimensional such that the reactive flow is always directed straight towards the local equilibrium (see Fig.~\ref{fig:2cMcRD_phase-space}a).
In contrast, in systems with more components, explicitly accounting for the relaxation towards local equilibria may be important to capture the salient features of the full dynamics.
One potential approach is to allow for small deviations from the local equilibria along the direction of the slowest decaying eigenvector. 
Moreover, local equilibria may become unstable, driving the concentrations away from them \cite{Halatek.Frey2018,Reichenbach.etal2007}. 
This qualitative change of the local reaction dynamics can have profound consequences on the dynamics of the spatially extended system, as was studied in detail in Refs.~\cite{Halatek.Frey2018,Brauns.etal2020a}.
There, it was found that destabilization of the local equilibria gives rise to chaos near the onset of pattern formation.

Even if a systematic reduction in terms of a (generalized) LQSSA is not possible, visualizing the trajectories from full numerical simulations in control space can be a powerful tool to gain insight into the underlying mechanisms \cite{Halatek.etal2018,Brauns.etal2020a}.

\begin{acknowledgments}
We thank Henrik Weyer for his critical reading of the manuscript. This work was funded by the Deutsche Forschungsgemeinschaft (DFG, German Research Foundation) -- Project-ID 201269156 -- Collaborative Research Center (SFB) 1032 -- Project B2.
\end{acknowledgments}



\appendix

\section{Relating diffusive exchange rates to diffusion constants} \label{app:LSA-continuous}

The diffusive exchange rates $\mathcal{D}_\alpha$ can be related to the diffusion constants $D_\alpha$ in a spatially continuous system in two alternative ways. First, a finite volume approximation of the Laplace operator on a line with reflective boundary conditions yields
\begin{equation} 
\label{eq:exchange-rates-FV}
	\mathcal{D}^\mathrm{(FV)}_\alpha
	= \frac{4}{L^2} D_\alpha.
\end{equation}
Second, we can choose the exchange rates such that the linearization of Eq.~\eqref{eq:two-compartment-dyn} for an antisymmetric perturbation $\uvec_{1,2}^{} = \uvec^* \pm \delta \uvec$ is identical to the linearization of a spatially continuous MCRD system for a Fourier mode ${\sim} \cos{qx}$ with $q  = \pi / L$:
\begin{equation} 
\label{eq:exchange-rates-LSA}
	\mathcal{D}^\mathrm{(LSA)}_\alpha
	= \frac{\pi^2}{2 L^2} D_\alpha.
\end{equation}
The factor $2$ in the denominator originates from the linearization of the exchange terms in Eq.~\eqref{eq:two-compartment-dyn} for the antisymmetric mode where any perturbation in compartment 1 is balanced by an equal and opposite perturbation in compartment 2. For symmetric perturbations $\uvec_{1,2}^{} = \uvec^* + \delta \uvec$, corresponding to homogeneous perturbations of the continuous system, the exchange term in Eq.~\eqref{eq:two-compartment-dyn} cancels.
For the exchange rates Eq.~\eqref{eq:exchange-rates-LSA}, the small amplitude dynamics of antisymmetric perturbations of the two-compartment system exactly represent the linearized dynamics of a single mode $q = \pi/L$ in the spatially continuous system, and one can use the system size $L$ as a bifurcation parameter to sample the whole dispersion relation $\sigma(q {\,=\,} \pi/L)$.

The two options above differ by a factor $\mathcal{D}^\mathrm{(LSA)}_\alpha/\mathcal{D}^\mathrm{(FV)}_\alpha = \pi^2/8 \approx 1.23$. This can be interpreted as an effective rescaling of the system size $L$ by a factor $\pi/(2\sqrt{2}) \approx 1.11$ due to the finite difference discretization. Throughout this study, we used the exchange rate defined by Eq.~\eqref{eq:exchange-rates-LSA}.

\section{Min system: Geometry reduction, parameter choice, numerical simulations and phase diagram}
\label{app:Min}

\subsection{Reduction from three-dimensional spherocylinder to two-comparmetment system}
\label{app:3D-to-compartments}

We model the three-dimensional cell geometry as a spherocylinder of length $L = \SI{3}{\micro m}$ and radius $R = \SI{0.5}{\micro m}$.
The surface of this spherocylinder represents the cell membrane and is the domain for the protein densities $\md$, $\mde$, while its three-dimensional bulk is the domain of the cytosolic components $\cDT, \cDD$ and $\cE$. Reactive boundary conditions at the surface account for attachment and detachment of proteins at the membrane. The mathematical implementation of the Min-skeleton model in this three-dimensional bulk-surface coupled setting can be found in \cite{Brauns.etal2020a}.

To reduce this geometry to the two-compartment system, we cut the spherocylinder at midplane and assume that the cytosol and membrane in both halves are well mixed. That is, we only account for concentration differences between the two cell halves which serve as a proxy for the concentration gradients along the cell. 
Moreover, we express the cytosol concentrations in units of surface density, $\hat{c} = \zeta c$, where $\zeta$ is the ratio of cytosolic bulk volume to membrane area (short bulk-surface ratio). This allows us to collect all concentrations in a vector that does not mix units. Substituting $c \to \hat{c}/\zeta$, all reaction rates for reaction terms involving a cytosol concentration are rescaled by the bulk-surface ratio: $\hat{k} = k/\zeta$. In the following, we drop the hats.

The bulk-surface ratio of a spherocylinder is given by
\begin{equation}
    \zeta = \frac{\pi R^2 L + 4 \pi R^3/3}{2 \pi R L + 4 \pi R^2}
          = \frac{R L + 4 R^2/3}{L + 4 R},
\end{equation}
which, with $L \approx \SI{3}{\micro m}$ and $R \approx \SI{3}{\micro m}$ for \emph{E.~coli}, gives $\zeta \approx 0.23$.

\begin{figure}
	\centering
	\includegraphics{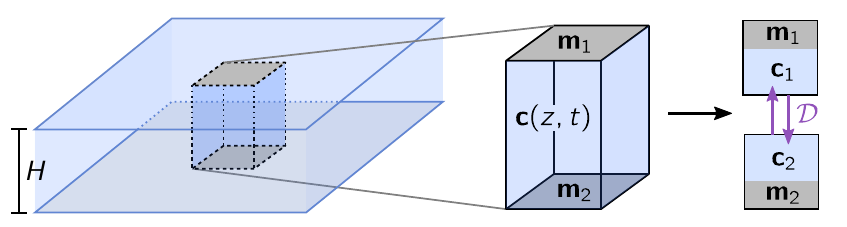}
	\caption{
	Illustration of an \textit{in vitro} setup using a flat microchamber with two membrane surfaces (gray planes) on top and bottom of a bulk volume \cite{Brauns.etal2020a}. An individual column of that system, comprising two membrane patches and the bulk volume in-between them can be pictured as an analog to the cell geometry, where the two membrane patches correspond to the cell poles. 
	The analogous approximation by two compartments, as shown on the right, is valid as long as the vertical bulk gradient is approximately linear.
	Comparing to Fig.~\ref{fig:motivation}a, the analogy between pole-to-pole oscillations in cells and vertical membrane-to-membrane oscillations in microchambers becomes immediately evident.
	}
	\label{fig:Min-in-vitro-reduction}
\end{figure}

For the \textit{in vitro} setup using flat microchambers whose top and bottom surfaces are covered by lipid bilayers that mimic the cell membrane, the bulk-surface ratio is simply $H/2$, where $H$ is the height of the microchamber; see Fig.~\ref{fig:Min-in-vitro-reduction}. With respect to this microchamber geometry, the two-compartment system represents a single, laterally isolated cytosol column with two membrane patches at its top and bottom. Only vertical gradients in the cytosol on the scale of the microchamber height are accounted for by the two compartments.

\begin{figure*}
	\centering
	\includegraphics{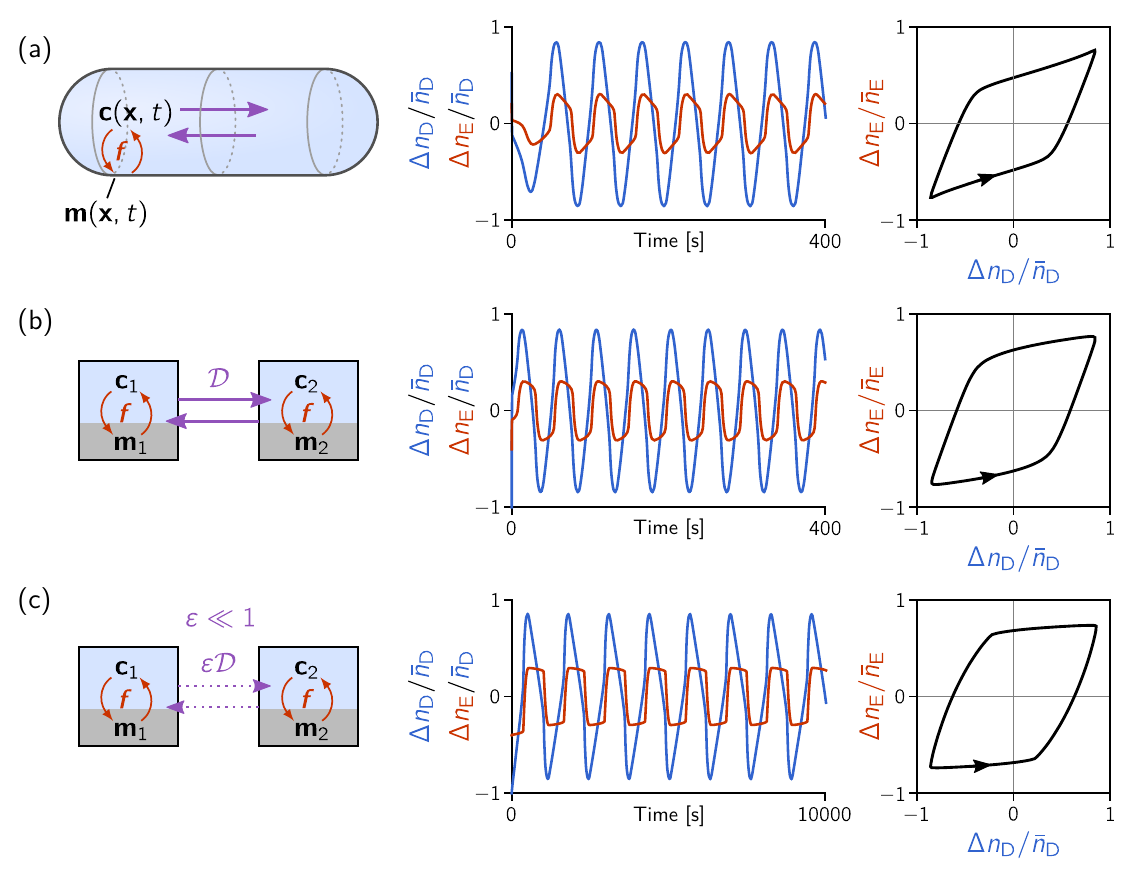}
	\caption{%
	Comparison of the Min-protein dynamics in the full 3D geometry of an \textit{E.\ coli} cell (a) to the two-compartment system (b,c) representing the two cell-halves (poles); see Sec.~\ref{sec:Min} for details.
	Nonlinear reactions ($f$, red arrows) account for cycling between membrane-bound and cytosolic states (concentrations $m$ and $c$). Diffusive exchange is indicated by purple arrows. 
	Time traces (center) and phase-space trajectories (right) of the redistributed masses $\dnDE$ between the two cell-halves/compartments show good qualitative agreement between the full 3D simulation and the two-compartment system.
	Importantly, setting the diffusive exchange rates to a much slower time scale ($\mathcal{D} \to \varepsilon \mathcal{D}$, here $\varepsilon = 10^{-2}$) does not qualitatively alter the pole-to-pole oscillation (c).
	}
	\label{fig:Min-systematic-reduction}
\end{figure*}

\subsection{Parameter choice}

\begin{table}[t]
    \caption{Parameters for the Min-skeleton model adapted from \cite{Halatek.Frey2012}. $\zeta$ is the bulk-surface ratio that appears because we express cytosol concentrations in units of surface density $\si{\micro m^{-2}}$, as explained in the text (Appendix.~\ref{app:3D-to-compartments}).}
    \label{tab:Min-params}
    \begin{ruledtabular}
    \renewcommand\arraystretch{1.2}
    \begin{tabular}{cll}
        Parameter & Value & Unit \\
    \hline
        $\DD$ & 16 & \si{\micro m^2.s^{-1}} \\
        $\DE$ & 10 & \si{\micro m^2.s^{-1}} \\
        $\Dd$ & 0.013 & \si{\micro m^2.s^{-1}} \\
        $\Dde$ & 0.013 & \si{\micro m^2.s^{-1}} \\
        $\nD$ & 5.0 & \si{\micro m^{-2}} \\
        $\nE$ & 2.0 & \si{\micro m^{-2}} \\
        $\lambda$ & 6.0 & \si{s^{-1}} \\
        $\kD$ & $0.1/\zeta$ & \si{s^{-1}} \\
        $\kdD$ & 0.$108/\zeta$ & \si{\micro m^2.s^{-1}} \\
        $\kdE$ & $0.435/\zeta$ & \si{\micro m^2.s^{-1}} \\
        $\kde$ & 0.4 & \si{s^{-1}} \\
    \end{tabular}
    \end{ruledtabular}
\end{table}

\begin{figure}
	\centering
	\includegraphics{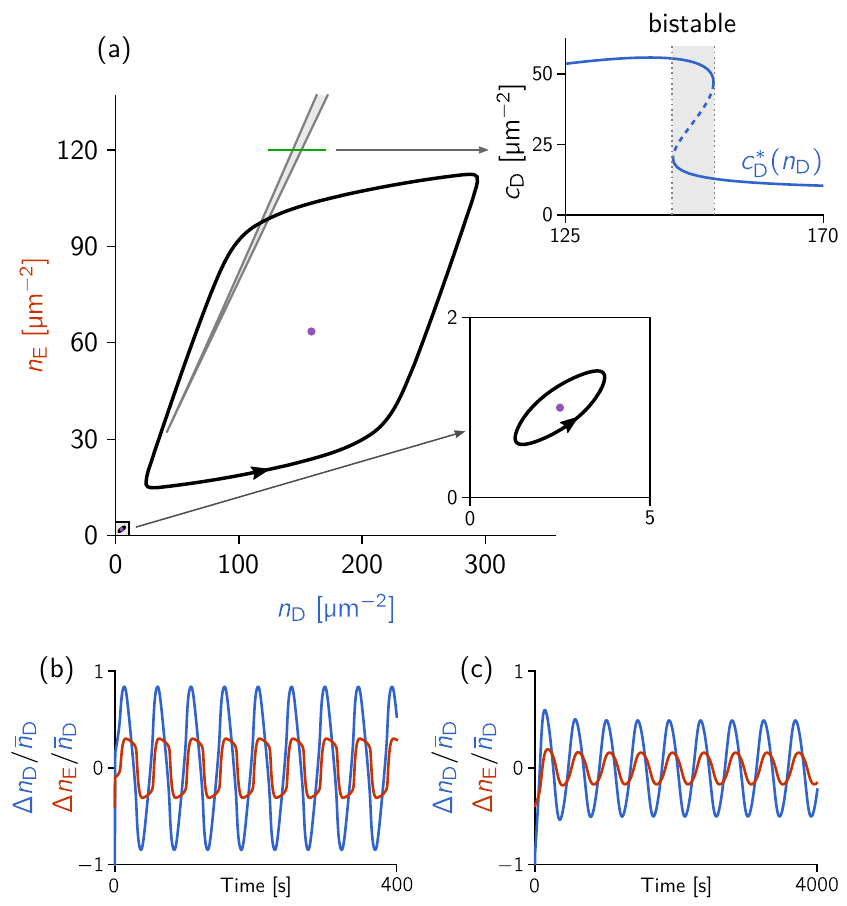}
	\caption{%
	Qualitative equivalence of the Min dynamics at physiological protein densities and with scaled down densities.
    (a) Parameter space of the protein densities $\nD, \nE$ showing the regime of bistable reaction kinetics in gray and trajectories from simulations of the two compartment model. The inset on the top right shows a curve of reactive equilibria (blue) in a slice through the bistable region at constant $\nE$. The inset on the bottom right shows a blow-up of the trajectory in the low density regime. 
    Purple dots mark the average masses.
    (b,c) Timetraces of the mass differences corresponding to the two trajectories in (a). Note the differently scaled time axes.
	}
	\label{fig:bistability-mass-scaling}
\end{figure}

For the physiological parameters from \cite{Halatek.Frey2012}, the densities enter a regime where the reaction kinetics is bistable (i.e.\ where there are two stable reactive equilibria for given local total densities, see Fig.~\ref{fig:bistability-mass-scaling}a). This ``local bistability'' does not change the dynamics of the spatially coupled system qualitatively. 
However, it complicates the analysis in terms of LQSSA to deal with the branch switching that happens when the dynamics leaves the locally bistable region: Upon passing the saddle-node bifurcations that delimit the bistable region, the concentrations jump to the remaining branch of stable equilibria. 
To avoid these technical subtleties, we reduce the total densities to values where the local system no longer becomes bistable (inset in Fig.~\ref{fig:bistability-mass-scaling}a). 
Because this also increases the minimal domain size for instability, we increase the domain length by a factor 8. The oscillation period increases due to this, but the limit cycle in control space does not change qualitatively (see Fig.~\ref{fig:bistability-mass-scaling}b,c). For the parameter values used here, see Table~\ref{tab:Min-params}.

\begin{figure}
    \centering
    \includegraphics{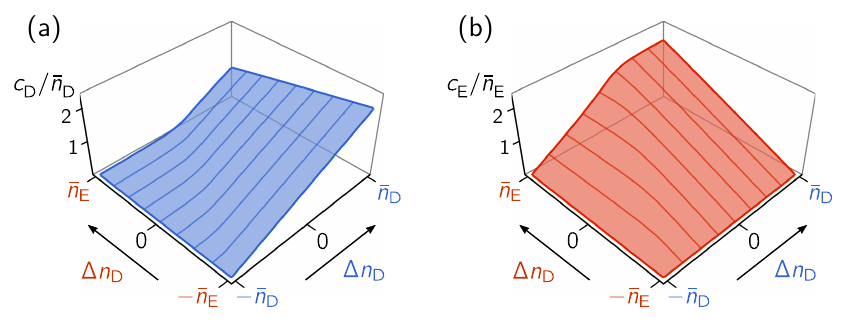}
    \caption{%
    Reactive equilibrium surfaces showing the membrane concentrations $\md$ and $\mde$.
    }
    \label{fig:Min-md-mde-surfaces}
\end{figure}

\subsection{Simulations on 1D domain}

In Figure.~\ref{fig:Min-phase-diagram}, we compare simulations of the two-compartment system to simulations in a spatially continuous domain (1D line) with no-flux boundary conditions. The dynamics in this domain is given by
\begin{equation}
    \partial_t \uvec(x,t) = \mathbf{D} \partial_x^2 \uvec + \fvec(\uvec),
\end{equation}
$\mathbf{D} = \diag(\{D_i\})$ is the diffusion matrix. (As in the two-compartment setting, the concentrations are measured in units of surface density, $\si{\micro m^{-2}}$. To convert the bulk concentrations to volume densities, they must be divided by the bulk-surface ratio $\zeta$.)

The reason that we do not perform the simulations in the three-dimensional cell geometry is that we are interested in the role of \emph{lateral} MinE transport, which we study by tuning the diffusion constants $\DE$ and $\Dde$. Bulk-surface coupling induces bulk-concentration gradients in the direction normal to the membrane. Those gradients control the flux onto and off the membrane (attachment--detachment dynamics). Hence, changing the cytosol diffusion constants in the bulk-surface coupled 3D system affects both transport and the reaction kinetics. Reducing the system to a 1D line geometry, which effectively amounts to neglecting vertical gradients, allows us to tune the cytosol diffusion constants to study the role of lateral mass transport alone.

\subsection{No instability for well-mixed cytosol}
\label{app:Min-well-mixed-cytosol}

In the limit $\DhD,\DhE \to \infty$, the cytosol is well mixed, i.e.\ $\cvec^{(1)} = \cvec^{(2)} = \cvec$. 
Defining
\begin{equation}
\begin{aligned}
    f_i^{(j)} = f_i(\md^{(j)},\mde^{(j)},\cDD,\cDT,\cE), \\ 
    i \in \{\mathrm{d,de,DD,DT,E}\}, \; j = 1,2,
\end{aligned}
\end{equation}
with $\fvec$ given by Eq.~\eqref{eq:Min-reactions}, the dynamics is governed by
\begin{subequations}
\begin{align}
    \partial_t \md^{(1,2)} &= \pm \Dhd \left(\md^{(2)} - \md^{(1)}\right) + f_\mathrm{d}^{(1,2)}, \\
    \partial_t \mde^{(1,2)} &= \pm \Dhde \left(\mde^{(2)} - \mde^{(1)}\right) + f_\mathrm{de}^{(1,2)}, \\
    \partial_t \cDD &= f_\mathrm{DD}^{(1)} + f_\mathrm{DD}^{(2)}, \\
    \partial_t \cDT &= f_\mathrm{DT}^{(1)} + f_\mathrm{DT}^{(2)}, \\
    \partial_t \cE &= f_\mathrm{E}^{(1)} + f_\mathrm{E}^{(2)}.
\end{align}
\end{subequations}

We now perform a linear stability analysis of the homogeneous steady states ($\mvec^{(1)} = \mvec^{(2)} = \mvec^*$, $\fvec(\uvec^*) = 0$) of these equations and show that they never exhibit a symmetry breaking instability.
Because of the parity symmetry, $1 \leftrightarrow 2$, of the homogeneous steady state, even and odd perturbations are decoupled. Even perturbations correspond to the stability against homogeneous perturbations. Odd perturbations correspond lateral stability, i.e.\ stability against spatially inhomogeneous perturbations. These are the relevant perturbations for pattern formation.
For odd perturbations, mass conservation of MinD and MinE enforces $\delta \cDD = -\delta \cDT$ and $\delta \cE = 0$. 
Thus, we obtain the eigenvalue problem
\begin{equation}
    \partial_t \begin{pmatrix} \delta \md \\ \delta \mde \\ \delta \cDT \end{pmatrix} =
    J_\mathrm{odd} \begin{pmatrix} \delta \md \\ \delta \mde \\ \delta \cDT \end{pmatrix}
\end{equation}
with the Jacobian
\begin{widetext}
\begin{equation}
	J_\mathrm{odd} = \begin{pmatrix}
		\kdD \cDT - \kdE \cE - 4 \Dhd & 0 & \kD + \kdD \md \\
		\kdE \cE & -\kde - 4 \Dhde & 0 \\
		0 & 0 & - 2\kD - 2 \kdD \md - 2 \lambda
	\end{pmatrix}.
\end{equation}
\end{widetext}
The eigenvalues of $J_\mathrm{odd}$ are 
\begin{equation}
\begin{aligned}
	\sigma_1 &= \kdD \cDT - \kdE \cE - 4 \Dhd, \\
	\sigma_2 &= -\kde - 4 \Dhde, \\
	\sigma_3 &= - 2\kD - 2 \kdD \md - 2 \lambda.
\end{aligned}
\end{equation}
One immediately sees that only the first eigenvalue, $\sigma_1$, may become positive.
A necessary condition for this is
\begin{equation} \label{eq:well-mixed-instability-condition}
	\kdD \cDT - \kdE \cE > 0.
\end{equation}
The Jacobian is evaluated at the homogeneous steady state where $\fvec(\uvec^*) = 0$. In particular,
\begin{equation}
	f_\mathrm{d} = (\kdD \cDT^* - \kdE \cE^*) \md^* + \kD \cD^* = 0,
\end{equation} 
which implies $\kdD \cDT^* < \kdE \cE^*$ for all steady states.
Therefore, the necessary condition for instability, Eq.~\eqref{eq:well-mixed-instability-condition}, is never fulfilled. 
In conclusion, the Min skeleton model with well-mixed cytosol cannot exhibit a lateral instability (instability against spatially inhomogeneous perturbations).
This result, derived in the two-compartment setting also holds in spatially continuous domains thanks to the correspondence between these two setting; see Sec.~\ref{app:LSA-continuous}.

\section{LQSSA}
\label{app:LQSSA}

\subsection{General setup and notation}
Consider a system with $N$ components, $\uvec = \{u_i\}_{i = 1..N}$, governed by local reactions $\fvec(\uvec)$ that conserve $M$ masses, $\nvec = \{n_\alpha\}_{\alpha = 1..M}$. 
The conserved masses are given in terms of the component vector via $n_\alpha = \mathbf{s}_\alpha^\mathsf{T} \uvec$ where $\mathbf{s}_\alpha$ are ``stoichiometric'' vectors fulfilling
$\mathbf{s}_\alpha^\mathsf{T} \fvec = 0$.
Denoting the diffusive exchange rates by the matrix $\mathfrak{D} = \diag \{\mathcal{D}_i\}$, the dynamics in LQSSA is given by
\begin{equation}
    \Delta n_\alpha = -\mathbf{s}_\alpha^\mathsf{T} \mathfrak{D} \Delta \uvec^*(\Delta \nvec),
\end{equation}
where the slaved concentration gradients $\Delta \uvec^*$ are defined in terms of the reactive equilibria as $\Delta \uvec^* = \uvec^*(\bar{\nvec} + \Delta \nvec) - \uvec^*(\bar{\nvec} - \Delta \nvec)$; cf.\ Eq.~\eqref{eq:scaffolded-gradients}.
The reactive equilibria $\uvec^*(\nvec)$ are defined by 
\begin{equation} \label{eq:loc-eq-general}
    0 \overset{!}{=} \mathbf{F}(\uvec^*; \nvec) = \begin{pmatrix} 
      \{\mathbf{s}_\alpha^\mathsf{T} \uvec^* - n_\alpha\}_\alpha \\
      \fvec(\uvec^*)
    \end{pmatrix}.
\end{equation}
The factor $s^\mathsf{T}_\alpha \, \mathfrak{D}$ determines the diffusive mass flux of species $\alpha$ that results from slaved concentration gradients. We now define the ``mass-redistribution potentials'' \cite{Brauns.etal2020b}
$\eta_\alpha := \mathbf{s}_\alpha^\mathsf{T} \mathfrak{D} \uvec$,
which allows us to write the mass-redistribution dynamics as 
\begin{equation}
    \partial_t \Delta n_\alpha = -\Delta \eta_\alpha^*(\Delta \nvec).
\end{equation}

\subsection{Linear stability analysis}
\label{app:LSA}

For small perturbations $\delta n$ around the homogeneous steady state $\Delta \nvec = 0$, the dynamics is given by 
\begin{align}
    \partial_t \delta n_\alpha &= 
    -2 \sum_\beta \partial_{n_\beta} \eta_\alpha^*|_{\bar{\nvec}} \, \delta {n_\beta}, \nonumber \\
    &= \sum_\beta \mathcal{J}_{\alpha\beta} \, \delta {n_\beta},
\end{align}
where, in the second line, we introduced the mass-redistribution Jacobian
\begin{equation}
    \mathcal{J}_{\alpha\beta} := - 2 \partial_{n_\beta} \eta_\alpha^*|_{\bar{\nvec}} = - 2 \mathbf{s}_\alpha^\mathsf{T} \, \mathfrak{D} \, (\partial_{n_\beta}\uvec^*|_{\bar{\nvec}}).
\end{equation}
The eigenvalues of this Jacobian determine the stability of the homogeneous steady state in LQSSA.

Before we continue to calculate the derivatives $\partial_{n_\beta} \eta_\alpha^*$ in terms of the linearized reaction kinetics, let us take moment to interpret the Jacobian $\mathcal{J}$. In the case of one conserved mass $n$, we have the $1{\times}1$ Jacobian $\mathcal{J} = -2\partial_n \eta*|_n$. Hence, we recover the nullcline-slope criterion for lateral instability $\partial_n \eta*|_n < 0$ (cf.\ Eq.~(27) in Ref.~\cite{Brauns.etal2020b}).
For more than one conserved mass, the entries of $\mathcal{J}$ are the slopes of the nullcline (hyper-)surfaces $\eta*(\nvec)$ along the directions of the conserved masses in parameter space. The eigenvalue problem for $\mathcal{J}$ can therefore be interpreted as a generalized slope criterion.

To find the nullcline slopes $\partial_{n_\beta} \eta_\alpha^*$, we take the derivative of the defining equation for the reactive equilibria Eq.~\eqref{eq:loc-eq-general} with respect to $n_\alpha$ which gives (implicit function theorem) 
\begin{equation}
    \partial_{n_\alpha} \uvec^* = -(D_\uvec \mathbf{F}|_{\uvec^*})^{-1}  \partial_{n_\alpha} \mathbf{F} = (D_\uvec \mathbf{F}|_{\uvec^*})^{-1} \mathbf{e}_\alpha,
\end{equation}
where $\mathbf{e}_\alpha$ is the unit vector with entry 1 in the $\alpha$th component.
Substituting this in Jacobian yields
\begin{equation}
    \mathcal{J}_{\alpha\beta} = -2 \mathbf{s}^\mathsf{T}_\alpha \, \mathfrak{D} \, (D_\uvec \mathbf{F}|_{\uvec^*(\bar{\nvec})})^{-1} \mathbf{e}_\beta.
\end{equation}
This can easily be implemented numerically to obtain the Jacobian and calculate its eigenvalues.

\paragraph{Equivalence to perturbation theory in long-wavelength limit.}
The Jacobian derived above for the two-compartement system in LQSSA can also be obtained by a long-wavelength perturbation theory for linear stability analysis on a continuous domain. To see why this is, consider the Jacobian on a spatially continuous domain
\begin{equation}
    J_q = D_\uvec \fvec |_{\bar{\nvec}} - q^2 \mathbf{D},
\end{equation}
where $\mathbf{D} = \diag(\{D_i\})$ is the diffusion matrix, and $q$ is the wavenumber (i.e.\ $-q^2$ are the eigenvalues of the Laplace operator). In the long wavelength limit $q \to 0$, we can find the eigenvalues of $J_q$ by solving the degenerate perturbation problem with $q^2$ as perturbation parameter. We are interested in the eigenvalue branches that emanate from $0$ at $q=0$, corresponding to the conservation laws.
The associated left eigenvectors, (spanning the left nullspace of $D_\uvec \fvec$) are the ``stochiometric vectors'' $\mathbf{s}_\alpha^\mathsf{T}$.
The right eigenvectors of $D_\uvec \fvec$ associated to the eigenvalue $0$ are $\partial_{n_
\alpha} \uvec^*$. This follows immediately from the defining equation $\fvec(\uvec^*) = 0$ by taking the derivative w.r.to $n_\alpha$ and using that $\fvec$ does not explicitly depend on $n_\alpha$.
The first order perturbation of the degenerate $0$ eigenvalues is given by eigenvalues of the matrix
\begin{equation}
    M^{(1)}_{\alpha\beta} = -\mathbf{s}_\alpha^\mathsf{T} \, \mathbf{D} \, (\partial_{n_\beta}\uvec^*),
\end{equation}
where we used $\mathbf{s}_\alpha^\mathsf{T} \partial_{n_\beta}\uvec^* = \partial_{n_\beta} (\mathbf{s}_\alpha^\mathsf{T} \uvec^*) = \partial_{n_\beta} n_\alpha = \delta_{\alpha\beta}$.
Substituting the diffusion matrix $\mathbf{D}$ by the exchange rate matrix via Eq.~\eqref{eq:exchange-rates-LSA} yields the desired result
\begin{equation}
    \mathcal{J}_{\alpha\beta} = q^2 M^{(1)}_{\alpha\beta}.
\end{equation}

\paragraph{Example: Min system.} For the Min system as defined in the main text we have
\begin{equation}
\begin{gathered}
    \uvec = (\md,\mde,\cDT,\cDD,\cE)^\mathsf{T}, \\
    \mathbf{s}_\mathrm{D}^\mathsf{T} = (1,1,1,1,0), \quad
    \mathbf{s}_\mathrm{E}^\mathsf{T} = (0,1,0,0,1), \\
    \mathbf{F} = \left(
      \mathbf{s}_\mathrm{D}^\mathsf{T} \uvec - n_\mathrm{D},\,
      \mathbf{s}_\mathrm{E}^\mathsf{T} \uvec - n_\mathrm{E},\,
      f_\mathrm{d}(\uvec), f_\mathrm{de}(\uvec), f_\mathrm{DD}(\uvec)
    \right), \\
    \mathfrak{D} = \diag(\Dhd,\Dhde,\DhD,\DhD,\DhE)\vphantom{^\mathsf{T}},
\end{gathered}
\end{equation}
where $\fvec$ is given in Eq.~\eqref{eq:Min-reactions}. Note that we eliminated to components from $\fvec$ because the system would otherwise be overdetermined owing to the two conserved masses.
This gives the derivative matrix
\begin{widetext}
\begin{equation}
    D_\uvec \mathbf{F} = \begin{pmatrix}
      1 & 1 & 1 & 1 & 0 \\
      0 & 1 & 0 & 0 & 0 \\
      \kdD \cDT - \kdE \cE & 0 & \kD + \kdD \md & 0 & -\kdE \md \\
      \kdE \cE & -\kde & 0 & 0 & \kdE \md \\
      0 & \kde & 0 & -\lambda & 0 \\
      - \kdE \cE & \kde & 0 & 0 & -\kdE \md
    \end{pmatrix}.
\end{equation}
\end{widetext}
Note that the first to rows are simply $\mathbf{s}_\mathrm{D}^\mathsf{T}$ and $\mathbf{s}_\mathrm{E}^\mathsf{T}$, which follows immediately from the definition of $\mathbf{F}$.

\paragraph{Inhomogeneous (asymmetric) steady states.}

The derivation presented above for homogeneous steady sates can be generalized to inhomogeneous steady states $\Delta\tilde{\nvec}$ defined by $\Delta \uvec^*(\Delta\tilde{\nvec}) = 0$. The resulting Jacobian reads
\begin{equation}
    \tilde{\mathcal{J}}_{\alpha\beta} = \mathbf{s}^\mathsf{T}_\alpha \, \mathfrak{D} \cdot \left[
        (D_\uvec \mathbf{F}|_{\bar{\nvec}+\Delta\tilde{\nvec}})^{-1} +
        (D_\uvec \mathbf{F}|_{\bar{\nvec}-\Delta\tilde{\nvec}})^{-1} 
    \right] \mathbf{e}_\beta.
\end{equation}

\section{PAR and Cdc42 models}
\label{app:PAR-and-Cdc42}

\subsection{PAR polarity model}

\begingroup
\newcommand\ma{m_\mathrm{a}}
\renewcommand\mp{m_\mathrm{p}}
\newcommand\cA{c_\mathrm{A}}
\newcommand\cP{c_\mathrm{P}}
\renewcommand\nA{n_\mathrm{A}}
\newcommand\nP{n_\mathrm{P}}

\newcommand\ka{k_\mathrm{a}}
\renewcommand\kA{k_\mathrm{A}}
\newcommand\kp{k_\mathrm{p}}
\newcommand\kP{k_\mathrm{P}}
\newcommand\kpa{k_\mathrm{pa}}
\newcommand\kap{k_\mathrm{ap}}

We adopt the model introduced in \cite{Goehring.etal2011} and further analyzed in \cite{Trong.etal2014} which accounts for the membrane-bound and cytosolic concentrations of aPARs and pPARs $\uvec = (\ma,\mp,\cA,\cP)$ with the reaction kinetics
\begingroup \renewcommand\arraystretch{1.25}
\begin{equation}
    \fvec = \begin{pmatrix} 
      \kA \cA - \ka \ma - \kap \mp^2 \ma \\
      \kP \cP - \kp \mp - \kpa \ma^2 \mp \\
      -\kA \cA + \ka \ma + \kap \mp^2 \ma \\
      -\kP \cP + \kp \mp + \kpa \ma^2 \mp
    \end{pmatrix}.
\end{equation}
\endgroup
These reactions conserve the total densities of aPARs $\nA = \ma + \cA$ and pPARs $\nP = \mp + \cP$, respectively. Since the reaction network is symmetric under the exchange A$\leftrightarrow$P, we use reaction rates that also respect this symmetry for simplicity \cite{Trong.etal2014}.
The diffusion matrix reads $\mathfrak{D} = 4/L^2 \diag(D_m,D_m,D_c,D_c)$, where $L \approx \SI{15}{\micro m}$ is the long half-axis of the ellipsoidal cells. For the model parameters, see Table~\ref{tab:PAR-params}. Note that in LQSSA, this length only contributes to the overall time scale but does not affect the phase portrait structure.

\begin{table}
    \caption{Parameters for the PAR model adapted from \cite{Trong.etal2014}.}
    \label{tab:PAR-params}
    \begin{ruledtabular}
    \renewcommand\arraystretch{1.2}
    \begin{tabular}{cll}
        Parameter & Value & Unit \\
    \hline
        $D_c$ & 10 & \si{\micro m^2.s^{-1}} \\
        $D_m$ & 0.1 & \si{\micro m^2.s^{-1}} \\
        $\nA$ & 3.0 & \si{\micro m^{-2}} \\
        $\nP$ & 3.0 & \si{\micro m^{-2}} \\
        $\ka$ & 0.1 & \si{s^{-1}} \\
        $\kA$ & 1.0 & \si{s^{-1}} \\
        $\kap$ & 0.1 & \si{\micro m^2.s^{-1}} \\
        $\kp$ & 0.1 & \si{s^{-1}} \\
        $\kP$ & 1.0 & \si{s^{-1}} \\
        $\kpa$ & 0.1 & \si{\micro m^2.s^{-1}} \\
    \end{tabular}
    \end{ruledtabular}
\end{table}
\endgroup

\subsection{Cdc42 polarity model}

\newcommand\mt{m_\mathrm{t}} 
\newcommand\mb{m_\mathrm{b}}
\newcommand\cB{c_\mathrm{B}}
\newcommand\nB{n_\mathrm{B}}

\newcommand\ktd{k_\mathrm{td}}
\newcommand\kd{k_\mathrm{d}}
\newcommand\kdt{k_\mathrm{dt}} 
\newcommand\kbd{k_\mathrm{bd}} 
\newcommand\kbD{k_\mathrm{bD}} 
\newcommand\kB{k_\mathrm{B}} 
\newcommand\kb{k_\mathrm{b}} 
\newcommand\ktB{k_\mathrm{tB}} 

We use a simplified form of the quantitative model proposed in \cite{Klunder.etal2013}. This model describes the dynamics of the GTPase Cdc42, its guanine nucleotide exchange factor (GEF) Cdc24 and the scaffold protein Bem1. The critical feedback loop is constituted by mutual recruitment between membrane-bound Cdc42-GTP and Bem1-GEF complexes. While the full model accounts for Bem1 and GEF separately, we lump these species into a complex species here. This retains the salient features of the model, in particular the mutual recruitment mechanism.

The variables of this simplified model are $\uvec = (\mt, \md, \mb, \cD, \cB)$, accounting, respectively, for membrane-bound Cdc42-GTP, Cdc42-GTP and Bem1-GEF complexes as well as cytosolic Cdc42-GDP and Bem1-GEF complexes.
The reaction kinetics, describing attachment and detachment of Cdc42 at the membrane, hydrolysis and nucleotide exchange of Cdc42 of membrane-bound Cdc42, as well recruitment of Bem1-GEF complexes to the membrane by Cdc42-GTP are given by
\begin{equation}
    \fvec = \begin{pmatrix} 
      \kdt \md + \kbd \mb \md + \kbD \mb \cD - \ktd \mt \\
      \kD \cD + \ktd \mt - (\kd + \kdt + \kbd \mb) \md \\
      \kB \cB + \ktB \mt \cB - \kb \mb \\
      - \kD \cD - \kbD \mb \cD + \kd \md \\
      - \kB \cB - \ktB \mt \cB + \kb \mb
    \end{pmatrix}.
\end{equation}
These reactions conserve the total densities of Cdc42 $\nD = \mt + \md + \cD$ and Bem1-GEF complexes  $\nB = \mb + \cB$, respectively.

\begin{table}
    \caption{Parameters for the Cdc42 model adapted from \cite{Klunder.etal2013}. $\zeta = R/3$ is the bulk-surface ratio of the spherical cell.}
    \label{tab:Cdc42-params}
    \begin{ruledtabular}
    \renewcommand\arraystretch{1.2}
    \begin{tabular}{cll}
        Parameter & Value & Unit \\
    \hline
        $D_c$ & 10 & \si{\micro m^2.s^{-1}} \\
        $D_m$ & 0.01 & \si{\micro m^2.s^{-1}} \\
        $R$ & 4.0 & \si{\micro m} \\
        $\nD$ & $3000 / (4 \pi R^2)$ & \si{\micro m^{-2}} \\
        $\nB$ & $6500 / (4 \pi R^2)$ & \si{\micro m^{-2}} \\
        $\kbd$ & 0.2 & \si{\micro m^2.s^{-1}} \\
        $\ktd$ & 1.0 & \si{s^{-1}} \\
        $\kdt$ & 0.002 & \si{s^{-1}} \\
        $\kbD$ & $0.266 / \zeta$ & \si{\micro m^2.s^{-1}} \\
        $\kd$ & 1.0 & \si{s^{-1}} \\
        $\kD$ & $0.28 / \zeta$ & \si{s^{-1}} \\
        $\kB$ & $0.001 / \zeta$ & \si{\micro m^2.s^{-1}} \\
        $\ktB$ & $0.009 / \zeta$ & \si{\micro m^2.s^{-1}} \\
        $\kb$ & 0.35 & \si{s^{-1}} \\
    \end{tabular}
    \end{ruledtabular}
\end{table}

The parameter values, given in Table~\ref{tab:Cdc42-params} are adapted from \cite{Klunder.etal2013}.
The values of $\kb$, $\kB$ and $\ktB$ are chosen to account for the lumped Bem1-GEF complexes species.

%

\end{document}